# Rapid Generation of Optimal Generalized Monkhorst-Pack Grids

*Yunzhe Wang[1], Pandu Wisesa[1], Adarsh Balasubramanian[1], Shyam Dwaraknath[2] and Tim Mueller[1],\**

[1] Department of Materials Science and Engineering, Johns Hopkins University, Baltimore, Maryland 21218, USA

[2] Lawrence Berkeley National Laboratory, Berkeley, California, 94720, USA

**Abstract**

Computational modeling of the properties of crystalline materials has become an increasingly important aspect of materials research, consuming hundreds of millions of CPU-hours at scientific computing centers around the world each year, if not more. A routine operation in such calculations is the evaluation of integrals over the Brillouin zone. We have previously demonstrated that performing such integrals using generalized Monkhorst-Pack $k$-point grids can roughly double the speed of these calculations relative to the widely-used traditional Monkhorst-Pack grids. However the generation of optimal generalized Monkhorst-Pack grids is not implemented in most software packages due to the computational cost and difficulty of identifying the best grids. To address this problem, we present new algorithms that allow rapid generation of optimal generalized Monkhorst-Pack grids on the fly. We demonstrate that the grids generated by these algorithms are on average significantly more efficient than those generated using existing algorithms across a range of grid densities. For grids that correspond to a real-space supercell with at least 50 angstroms between lattice points, which is sufficient to converge density functional theory calculations within 1 meV/atom for nearly all materials, our algorithm finds optimized grids in an average of 0.19 seconds on a single processing core. To facilitate the widespread adoption of this approach, we present new open-source tools including a library designed for integration with third-party software packages.

1. Introduction

Computational materials research has become increasingly vital in probing the properties of crystalline materials, especially in screening materials at a large scale to accelerate material discoveries for a wide range of applications. A routine operation for such calculations across a variety of computational methods is the evaluation of integrals over the Brillouin zone, which can be approximated by discretely sampling the Brillouin Zone at a set of points known as *k*-points. Many popular computational materials simulation packages generate *k*-points using the traditional Monkhorst-Pack scheme,[1] which creates regular *k*-point grids with lattice vectors that are integer fractions of a particular set of reciprocal lattice vectors. We have demonstrated in our previous work that the number of symmetrically irreducible *k*-points, and hence the computational cost of most methods that rely on *k*-point sampling, can be reduced by roughly a factor of two by generalizing the Monkhorst Pack scheme so that the grids do not need to be aligned with a particular set of reciprocal lattice vectors and selecting the optimal generalized grid.[2] The benefits of using generalized grids can be understood by considering that the set of generalized *k*-point grids is a superset of traditional Monkhorst Pack grids, providing far more options for selecting the optimal grid. Other researchers have since found similar results.[3, 4]

Calculating the properties of crystalline materials consumes hundreds of millions of CPU-hours at supercomputing centers around the world each year, if not much more. (A single high-throughput project, the Materials Project, spends more than 100 million CPU hours per year calculating the properties of crystalline materials.) Given that modern high-performance computing resources cost about US$ 0.0255 per CPU hour[1] or more,[5] we conservatively estimate

---

[1] The CPU price is the latest listed value for the standard AWS machine type a1.medium with 2GB memory.

that the use of generalized Monkhorst Pack grids in place of traditional grids has the potential to save researchers millions of U.S. dollars per year in computing costs.

Some of the ideas behind the generalized *k*-point grids had been proposed by Froyen and Moreno and Soler decades ago,[1, 6] but they have not been widely adopted primarily due to the computational challenge of identifying the best generalized grid for a given calculation. The main challenge is that the number of possible generalized *k*-point grids grows rapidly with the number of *k*-points in the grid (section 2 of the Supplementary Information), making it difficult to identify which grid is most efficient.[2, 7] For example, there are 54,156,102 regular grids that contain 4,000 *k*-points, a typical density for calculations on elemental metals. Identifying the optimal grid requires identifying which among these candidates is expected to provide a sufficiently accurate estimate of the integral with the fewest symmetrically irreducible *k*-points. The problem is made more challenging by the fact that it is generally necessary to search over many different *k*-point densities to find the optimal grid.

In our previous work we addressed these problems by creating a free, internet-accessible *k*-point grid server, backed by a database of pre-calculated generalized grids, that rapidly returns an efficient grid (typically the most efficient grid) for a given calculation.[2] To date, this server has delivered more than half a million grids to users outside our research group. In the years since our previous work was published there has been increasing interest in the generation and use of generalized *k*-point grids[4, 8-29] and how they may be used in popular software packages.[28] Yet despite the increasing interest in the use of regularized grids, most common software packages do not yet implement an efficient method for identifying highly efficient generalized grids, due largely to the lack of publicly available algorithms and tools for doing so.

To enable more widespread use of generalized Monkhorst-Pack *k*-point grids and fully realize their potential for accelerating computational materials research, we have developed an open-source library for grid generation, kpLib, that is designed for integration with existing software packages without significantly increasing the size of their software distribution. This library is based on novel algorithms, described in this manuscript, that greatly accelerate grid generation. These algorithms include a method for significantly reducing the number of candidate superlattices to be evaluated by transforming the problem from an enumeration of 3D superlattices to an enumeration 2D superlattices with a finite set of allowed stackings. We have also developed an open-source standalone tool for generalized *k*-point grid generation, the *K*-Point Grid Generator. This tool has the same functionality as the *K*-Point Grid Server, but it can be used on computing nodes that do not have network access to the *K*-Point Grid Server. Additional algorithms for the *K*-Point Grid Generator and its implementation are described in detail in sections 3, 5 and 6.2 of the Supplementary Information.

To illustrate the performance of kpLib, we present benchmarks on structures randomly selected from the Inorganic Crystal Structure Database.[30] Our benchmarks demonstrate that at a grid density sufficient to converge calculated energies on nearly all crystalline materials within 1 meV / atom, kpLib identifies the optimal grid in less than half a second on average, and in under five seconds for grids that are eight times as dense. We further demonstrate that on average our algorithm finds grids with significantly fewer irreducible *k*-points than an alternative algorithm for generating generalized Monkhorst-Pack grids recently developed by Hart and co-workers.[31, 32]

In the following sections, a detailed explanation of the new algorithms is provided, and the implementation of kpLib is briefly discussed. Various benchmarks of the speed of the algorithms and quality of the resulting grids are then provided. Additional comparisons between kpLib and the *K*-Point Grid Generator, along with detailed descriptions of other algorithms used by these software packages, are provided in the Supplementary Information.

## 2. Algorithms

### 2.1 Background and notation

Monkhorst-Pack grids are used to approximate the value of an integral over the Brillouin zone by sampling reciprocal space on a regular grid of *k*-points, where the coordinates of the *k*-points are given by

$$\mathbf{k} = \frac{n_1}{m_1}\mathbf{b}_1 + \frac{n_2}{m_2}\mathbf{b}_2 + \frac{n_3}{m_3}\mathbf{b}_3 + \mathbf{s}, \quad (1)$$
$$n_1 = 0\ldots m_1 - 1,\ n_2 = 0\ldots m_2 - 1,\ n_3 = 0\ldots m_3 - 1$$

where $m_1$, $m_2$, and $m_3$ are positive integers, $\mathbf{b}_1$, $\mathbf{b}_2$ and $\mathbf{b}_3$ are reciprocal lattice vectors, and $\mathbf{s}$ represents a shift vector that moves the grid away from the origin (known as the Γ point in reciprocal space). There exists a mapping between each regular *k*-point grid and a real-space superlattice that defines the Born-von Karman boundary conditions for the periodicity of the wave functions.[33, 34] The superlattice corresponding to the *k*-point grid defined by equation (1) is given by

$$(\mathbf{g}_1, \mathbf{g}_2, \mathbf{g}_3)^T = \mathbf{M}(\mathbf{a}_1, \mathbf{a}_2, \mathbf{a}_3)^T \quad (2)$$

where $\mathbf{a}_1$, $\mathbf{a}_2$, and $\mathbf{a}_3$ represent the real-space primitive lattice vectors, $\mathbf{g}_1$, $\mathbf{g}_2$ and $\mathbf{g}_3$ represent the lattice vectors of the superlattice, and the transformation matrix $\mathbf{M}$ is equal to

$$\mathbf{M} = \begin{bmatrix} m_1 & 0 & 0 \\ 0 & m_2 & 0 \\ 0 & 0 & m_3 \end{bmatrix}. \quad (3)$$

The reciprocal primitive lattice vectors share an analogous relationship with those of the reciprocal superlattice. The reciprocal lattice vectors of a direct lattice are calculated by

$$[\mathbf{b}_1, \mathbf{b}_2, \mathbf{b}_3]^T = [\mathbf{a}_1, \mathbf{a}_2, \mathbf{a}_3]^{-1} \tag{4}$$

where the vectors share the same definition as in equations (1) and (2). Similarly, the primitive reciprocal lattice vectors of the superlattice can be obtained by

$$[\mathbf{d}_1, \mathbf{d}_2, \mathbf{d}_3]^T = [\mathbf{g}_1, \mathbf{g}_2, \mathbf{g}_3]^{-1} \tag{5}$$

where $\mathbf{d}_1$, $\mathbf{d}_2$, and $\mathbf{d}_3$ are the reciprocal lattice vectors corresponding to the direct superlattice. Substituting equations (4) and (5) into equation (2), the following relationship can be derived:

$$[\mathbf{b}_1, \mathbf{b}_2, \mathbf{b}_3]^T = \mathbf{M}^T [\mathbf{d}_1, \mathbf{d}_2, \mathbf{d}_3]^T. \tag{6}$$

The matrix multiplication order implies that the row vectors of the matrix $\mathbf{M}^T$ contain the coordinates of the vectors $\{\mathbf{b}_1, \mathbf{b}_2, \mathbf{b}_3\}$ in the basis of $\{\mathbf{d}_1, \mathbf{d}_2, \mathbf{d}_3\}$.

In terms of the matrix $\mathbf{M}$, equation (1) can be written as

$$\begin{aligned} \mathbf{k} &= (n_1, n_2, n_3) \left( \begin{bmatrix} m_1 & 0 & 0 \\ 0 & m_2 & 0 \\ 0 & 0 & m_3 \end{bmatrix}^{-1} \right)^T (\mathbf{b}_1, \mathbf{b}_2, \mathbf{b}_3)^T + \mathbf{s} \\ &= (n_1, n_2, n_3)(\mathbf{M}^{-1})^T (\mathbf{b}_1, \mathbf{b}_2, \mathbf{b}_3)^T + \mathbf{s} \\ &= (n_1, n_2, n_3)[\mathbf{d}_1, \mathbf{d}_2, \mathbf{d}_3]^T + \mathbf{s} \end{aligned} \tag{7}$$

Therefore, the set of vectors $\{\mathbf{d}_1, \mathbf{d}_2, \mathbf{d}_3\}$ are a generating basis of the *k*-point grid. As shown in equation (7), the traditional Monkhorst-Pack scheme uses a diagonal matrix $\mathbf{M}$, which is equivalent to the constraint that the *k*-point grids are aligned with the reciprocal lattice vectors. However Froyen has pointed out that this constraint is not necessary,[6] and we have previously demonstrated that much more efficient grids can be generated if the Monkhorst-Pack approach is generalized by relaxing this requirement.[2] The resulting generalized *k*-point grids, as shown by Moreno and Soler, can always be represented as standard Monkhorst-Pack grids provided a suitable set of reciprocal lattice vectors are chosen.[7] Mathematically, this is equivalent to perform a diagonal decomposition on the integer matrix $\mathbf{M}$ by unimodular matrices

$$\mathbf{M} = \mathbf{U}\mathbf{D}\mathbf{U}^{-1} \quad (8)$$

and transforming the reciprocal lattice vectors to an equivalent set by plugging it into equation (6):

$$[\mathbf{b}_1, \mathbf{b}_2, \mathbf{b}_3]^T = (\mathbf{U}\mathbf{D}\mathbf{U}^{-1})^T [\mathbf{d}_1, \mathbf{d}_2, \mathbf{d}_3]^T$$
$$(\mathbf{U}^T [\mathbf{b}_1, \mathbf{b}_2, \mathbf{b}_3]^T) = \mathbf{D}(\mathbf{U}^T [\mathbf{d}_1, \mathbf{d}_2, \mathbf{d}_3]^T) \quad (9)$$
$$[\mathbf{b}'_1, \mathbf{b}'_2, \mathbf{b}'_3]^T = \mathbf{D}[\mathbf{d}'_1, \mathbf{d}'_2, \mathbf{d}'_3]^T$$

where $\mathbf{b}'_1$, $\mathbf{b}'_2$, and $\mathbf{b}'_3$ are the reciprocal lattice vectors that diagonalize the generating matrix. Thus generalized Monkhorst Pack *k*-point grids can be used for all of the same types of calculations that traditional Monkhorst-Pack grids are used for.

Equations (2) and (7) demonstrate that the search for optimal generalized *k*-point grids can be accomplished by an iteration over real-space superlattices, specified by the matrix $\mathbf{M}$, and shift vectors, given by the vector $\mathbf{s}$. Since the quality of *k*-point grids are determined by the number of symmetrically irreducible *k*-points, all symmetries of structures should be preserved in the grids, which transfers to the requirements that the corresponding superlattices must also be symmetry-preserving. In the following discussion, we use the symbols $r_{lattice}$, $N_i$, and $N_T$ to represent, respectively, the minimum spacing between points on the a superlattice, the number of symmetrically irreducible *k*-points, and the number of total *k*-points in the Brillouin zone. $N_T$ is also then the number of primitive cells in a unit cell of the corresponding real-space superlattice (aka the "size" of the superlattice), and is given by the absolute value of the determinant of $\mathbf{M}$.

## 2.2 A New Algorithm for Dynamically Generating Generalized K-Point Grids

Although the benefits of using generalized *k*-point grids are well-established,[2-4, 24] they have not yet been widely implemented in common software packages due primarily to the challenge in implementing an algorithm for efficiently generating them. To address this problem and facilitate the generation of generalized *k*-point grids in common materials software packages, we have developed a novel algorithm for rapidly and dynamically identifying a highly efficient generalized *k*-point grid. Unlike our previous approach, this algorithm does not make use of a database, allowing us to implement it in a lightweight, open-source library designed to be integrated with third-party software packages. Although the lack of a database reduces the speed of grid generation (see section 4.1), we expect the optimized dynamic generation algorithm we present here to be sufficiently fast for most practical applications. We have also released a standalone open-source tool that provides additional functionality and makes use of a database, using algorithms described in section 3 of the supporting information.

The dynamic grid generation method starts with three parameters describing the input structure:

1. The real-space primitive lattice vectors, $\{\mathbf{a}_1, \mathbf{a}_2, \mathbf{a}_3\}$.

2. The real-space conventional lattice vectors, $\{\mathbf{c}_1, \mathbf{c}_2, \mathbf{c}_3\}$, where at least one of the vectors is orthogonal to the other two for all but triclinic systems.

3. The group of point symmetry operations, $\{\mathbf{R}\}$, that the *k*-point grid (and real-space superlattice) should preserve. These point symmetry operations can be generated by removing translation from all the operations in the real-space crystallographic space group, resulting in a symmorphic space group. If the system has time reversal symmetry, then the reciprocal-space band structure will have inversion symmetry even if the real-space crystal does not. In this case, inversion and any additional operators required to complete the group should be added if they are not already present.

The algorithm then searches for the *k*-point grid that minimizes $N_i$ while satisfying the following two constraints:

1. $r_{lattice}$ for the corresponding superlattice not smaller than $r_{min}$ (a value provided by the user),
2. $N_T$ is greater than or equal to $N_{min}$ (another value provided by the user).

We start by determining a lower bound for $N_T$, which we call, $N_{lower}$. It is the larger value of $N_{min}$ and the minimum size that any superlattice can have with while satisfying $r_{lattice} \geq r_{min}$:

$$N_{lower} = \max\left(N_{min}, \left\lfloor \frac{\sqrt{2}}{2} r_{min}^3 \Big/ V_p \right\rfloor\right) \qquad (10)$$

where $V_p$ is the volume of the primitive cell, $\frac{\sqrt{2}}{2} r_{min}^3$ is the volume of a unit cell in a face-centered cubic (fcc) lattice for which the distance between lattice points is $r_{min}$, $\lfloor x \rfloor$ is the floor operation that returns the largest integer no greater than the argument $x$. Equation (10) can be justified by considering that fcc structures maximize the packing density for rigid spheres[35] and thus $\frac{\sqrt{2}}{2} r_{min}^3$ is the minimum unit cell volume for a superlattice for which $r_{lattice}$ is at least $r_{min}$.

The search for optimal superlattices starts with lattices of size $N_{lower}$ and generates symmetry-preserving superlattices using an algorithm to be introduced in section 2.3. For each symmetry-preserving superlattice, the scheme checks whether $r_{lattice}$ is smaller than $r_{min}$ and discards it if it is. When the first superlattice for which $r_{lattice} \geq r_{min}$ is found, its corresponding *k*-point grid is kept as the initial "best grid", and the scheme can determine an upper limit for the search, $N_{upper}$:

$$N_{upper} = N_i \times N_{sym} \tag{11}$$

where $N_{sym}$ is the number of unique point symmetry operations for the system, as provided in the third input parameter listed above. Any superlattices with $N_T \geq N_{upper}$ would necessarily have more irreducible *k*-points than that of the initial best grid. If at some point a superlattice with $N_i$ smaller than that of the best known grid is found, the best grid is updated to this newly found one and the value of $N_{upper}$ is adjusted accordingly. When two *k*-point grids have the same $N_i$, the scheme favours the one with a larger $r_{lattice}$ in the corresponding superlattice. If $r_{lattice}$ of both superlattices also tie, the scheme chooses the one with a larger $N_T$. The search ends when the upper limit of the sizes of superlattices is reached. Figure 1 summarizes the steps of the scheme.

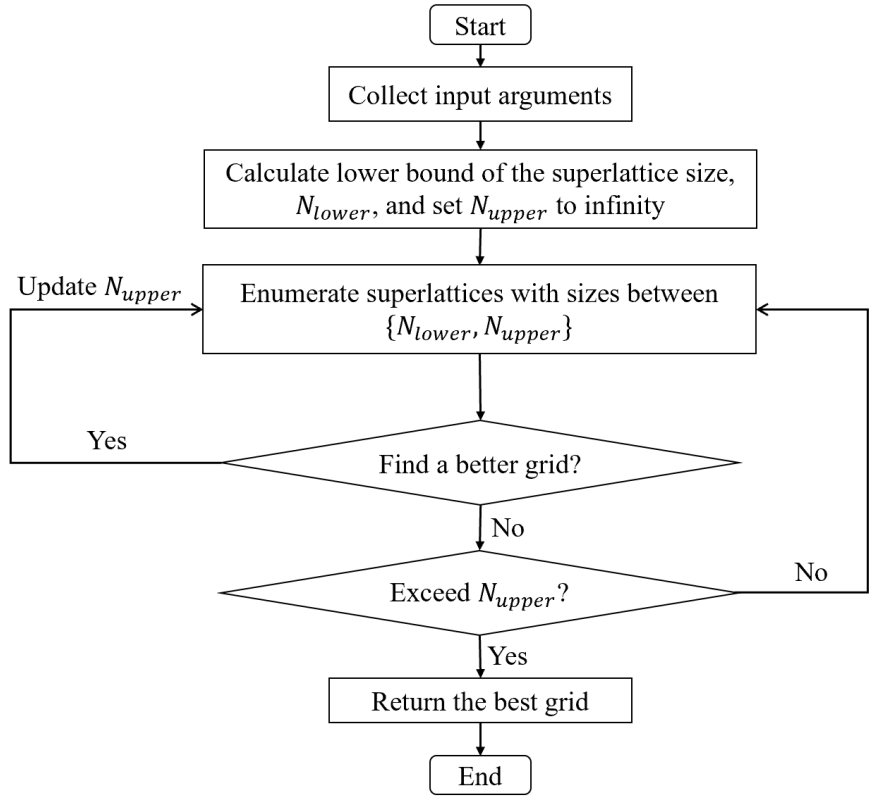

Figure 1. A diagram summarizes the workflow of the dynamic grid generation algorithm.

**2.3 Algorithms for Efficient Enumeration of Symmetry-Preserving Superlattices**

Enumeration of all symmetry-preserving superlattice is computationally expensive and has been identified as the main hurdle of applying generalized *k*-point grids in calculations of properties of crystalline materials.[2,7] Morgan et al. have presented an algorithm for accelerating the enumeration of symmetry-preserving lattices for a given lattice size by expressing the primitive lattice in Niggli-reduced form.[32] For each of the 44 distinct Niggli bases, they have determined symmetry-based constraints on the entries of **H** that can be used to reduce the number of possible lattices that must be considered. We have developed an approach that similarly iterates over symmetry-preserving lattices, with two key differences: it does not rely on Niggli reduction, which reduces the complexity of the code and increases the ease of implementation, and it is optimized for grid selection based on $r_{min}$, which has been shown to work well as a descriptor of *k*-point density both in theory[2] and in practice.[2,4] In our benchmarks, we demonstrate that the algorithms presented here generally return more efficient grids than the those generated using the method of Morgan et al.[32]

### 2.3.1 Hermite normal form and symmetry-preserving lattices

It is possible for two different matrices **M** to represent the same superlattice; i.e. the rows of each matrix could represent a different choice of vectors used to represent the lattice. For the purpose of enumerating over lattices we express the transformation matrix **M** in Hermite normal form, a triangular form which uniquely defines a superlattice.[36,37] We shall use **H** to represent the Hermite normal form of a general matrix **M**.

Efficient *k*-point grids will generally have symmetry-preserving lattices, which are invariant with respect to the symmetry operations of the system. Hermite normal form provides a convenient way to test whether a superlattice is symmetry-preserving by generating the Hermite normal forms for the original matrix **M** and all matrices generated by applying the symmetry operations of the system to $\mathbf{M}$. If all of the generated Hermite normal forms are the same, the lattice is symmetry-preserving.

### 2.3.2 Enumeration Algorithm for Crystal Systems Other than Triclinic

We start by considering systems that are not triclinic. For such systems at least one of the conventional lattice vectors must, by the symmetry of the system, be perpendicular to the other two. For simplicity, our only requirement is that such a vector be listed third, as $\mathbf{c}_3$.

The key to our approach is the recognition that for systems that are not triclinic, any regular three-dimensional lattice consists of layers of identical two-dimensional lattices that are normal to $\mathbf{c}_3$. Each two-dimensional lattice may be shifted from the one below it by a constant shift vector that is parallel to its lattice plane, and for symmetry-preserving lattices only a finite set of shift vectors are allowed. This decomposition helps quickly rule out superlattices that break symmetries without applying linear algebra to check them. For example, if there is a twofold rotational axis parallel to $\mathbf{c}_3$, then this axis may only pass through points in the two-dimensional lattice formed by linear combinations of half lattice vectors (Table 1). Any other shift would result in a lattice that is not symmetry preserving, as symmetry operations could transform lattice points to non-lattice points. Similarly, if there is a mirror plane perpendicular to $\mathbf{c}_3$, then either the mirror plane must be at the mid-point between two layers, in which case no shift is allowed, or it must pass through one of the layers, and again only the shifts shown in Table 1 are allowed. This concept is illustrated in two dimensions in Figure 2. Similar sets of shifts may be derived for three-fold rotational axes (Table 1).

A high-level summary of our algorithm for enumerating symmetry-preserving lattices is then as follows:

1. Determine all pairs of factors of the total lattice size. In each pair, the first factor represents the size of the supercell in each two-dimensional layer and the second represents the number of layers in each three-dimensional supercell.

2. For each pair of factors, enumerate all symmetry-preserving two-dimensional lattices (in Hermite normal form) with the required size.

3. Combine each two-dimensional lattice with each allowed shift to create a candidate three-dimensional lattice.

4. Verify that the three-dimensional lattice is symmetry-preserving.

| Crystal System | Shift vectors in the basis of $\{c_1, c_2\}$ in real space | Shift vectors of the $\Gamma$ point in the basis of $\{d_1, d_2, d_3\}$ as defined in equation (5) |
|---|---|---|
| Cubic, Tetragonal, Orthorhombic, Monoclinic | [0.0, 0.0], [0.0, 0.5], [0.5, 0.0], [0.5, 0.5] | [0.0,0.0,0.0], [0.0,0.0,0.5], [0.0,0.5,0.0], [0.5,0.0,0.0], [0.5,0.5,0.0], [0.5,0.0,0.5], [0.0,0.5,0.5], [0.5,0.5,0.5] |
| Hexagonal, Trigonal | [0.0, 0.0], [1/3, 0.0], [0.0, 1/3], [0.0, 2/3], [2/3, 0.0], [1/3, 1/3], [2/3, 2/3], [1/3, 2/3], [2/3, 1/3] | |

Table 1. Possible displacements of lattice planes in real space in 2 dimensions, and of the $\Gamma$ point in reciprocal space in 3 dimensions.

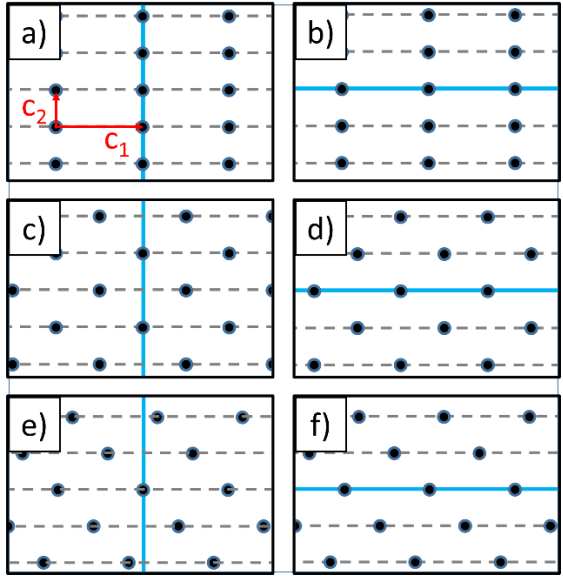

Figure 2. Two-dimensional examples of allowed and disallowed shifts. In all examples blue lines represent a mirror plane, black dots represent lattice points on real-space superlattice, and dashed lines show the different layers of lattice points that are orthogonal to $c_2$. a), b), c), and d) show allowed shifts in which the mirror plane transforms every lattice point to another lattice point. In a) and b) there is zero shift, and in c) and d) the shift is half the vector $c_1$. e) and f) show disallowed shifts.

This algorithm effectively reduces the problem of enumerating three-dimensional lattices to one of enumerating two-dimensional lattices, which significantly accelerates the search for symmetry-preserving lattices. Firstly, it drastically decreases the total number of 3-dimensional superlattices that need to be checked for symmetry preservation. Secondly, the symmetry groups in the 2-dimensional sublattice have fewer symmetry operations than the corresponding groups in 3 dimensions. Thirdly, a 2-dimensional matrix multiplication takes fewer elementary operations than a 3-dimensional one. We can even further accelerate the search by recognizing that if the number of layers is too small to satisfy the requirement that $r_{lattice} \geq r_{min}$, we can skip the enumeration of two-dimensional lattices and move on to the next set of factors. Similarly, if we ever determine that the $r_{lattice} < r_{min}$ for any two-dimensional layer, then we can stop evaluation of all lattices constructed from that layer and move onto the next two-dimensional lattice. We find that pre-screening the lattices for $r_{lattice}$ in this way significantly increases the speed of the algorithm when $r_{min}$ is the limiting factor, as demonstrated by the benchmarking results in section 7.2 of the Supplementary Information.

The steps of the algorithm are shown in detail by the pseudocode in Figure 3. The term "maxZDistance" at line 6 defines the maximum possible length of the shortest vector parallel with $\mathbf{c}_3$ that superlattices can have while satisfying $r_{lattice} \geq r_{min}$. The function "symmetryPreserving($\mathbf{H}$, {$\mathbf{R}$})" determines whether the set of symmetries is preserved in the given superlattice by checking the invariance of $\mathbf{H}$ after applying symmetries. Line 28 verifies that candidate lattices are superlattices of the primitive lattice after shifts in Table 1 are applied.

**Algorithm 1** Fast Enumeration of Symmetry-Preserving Superlattices

**Input:**
  $N_T$ - Size of superlattice. Also the total number of $k$-points in the grid.
  $\{c_1, c_2, c_3\}$ - Conventional lattice vectors.
  $\{R\}$ - The point symmetry group to be preserved in superlattices.
  $\{a_1, a_2, a_3\}$ - Primitive lattice vectors.
  $\{s\}$ - Array of possible 2D shift vectors.
  $r_{min}$ - Minimum periodic distance specified by users.

**Output:**
  $\{\{g_1, g_2, g_3\}\}$ - Array of symmetry-preserving superlattices with $r_{lattice} \geq r_{min}$.

**Initialization:**
  $\{\{g_1, g_2, g_3\}\}$ - Empty array of symmetry-preserving superlattices.
  $H$ - A $2 \times 2$ zero matrix.
  $M$ - A $3 \times 3$ zero matrix.

1: find the point symmetry group, $\{R^{2D}\}$, of the 2D sublattice $\{c_1, c_2\}$;
2: factorSets[][] ← sets of factorizations of $N_T$ into 2 integral numbers;
3: **for** factors[ ] **in** factorSets[ ][ ] **do**
4:     primLayerSpacing = $\|c_3\| \times V_p/V_c$;
5:     maxLayers = *isHexagonal*() ? 3 : 2;
6:     maxZDistance = factors[2] × primLayerSpacing × maxLayers;
7:     **if** maxZDistance < $r_{min}$ **then**
8:         continue;
9:     $\{\{g_1, g_2\}\}$ ← initialize an empty array of 2 vectors;
10:     factorSets2D[ ][ ] ← sets of factorization of factors[1] into 2 integral numbers;
11:     **for** factors2D[ ] **in** factorSets2D[ ][ ] **do**
12:         $H_{11}$ = factors2D[1], $H_{22}$ = factors2D[2]
13:         **for** $i = 0$ **to** factors2D[1] - 1 **do**
14:             $H_{21} = i$;
15:             **if** not *symmetryPreserving*($H, \{R^{2D}\}$) **then**
16:                 continue;
17:             $g_1 = H_{11}c_1$;
18:             $g_2 = H_{21}c_1 + H_{22}c_2$;
19:             add $\{g_1, g_2\}$ to the array of 2D symmetry-preserving superlattices, $\{\{g_1, g_2\}\}$;
20:     **for** $\{g_1, g_2\}$ **in** $\{\{g_1, g_2\}\}$ **do**
21:         **if** *getMinDistance*($g_1, g_2$) < $r_{min}$ **then**
22:             continue;
23:         **for** s **in** $\{s\}$ **do**
24:             $g_3 = [s_1, s_2] \cdot [g_1, g_2]^T + (\text{factors}[2]/(V_c/V_p)) \cdot c_3$;
25:             **if** *getMinDistance*($g_1, g_2, g_3$) < $r_{min}$ **then**
26:                 continue;
27:             $M = [g_1, g_2, g_3]^T \cdot \left([a_1, a_2, a_3]^T\right)^{-1}$;
28:             **if** $M$ contains non-integral elements **then**
29:                 continue;
30:             **if** not *symmetryPreserving*($M, \{R\}$) **then**
31:                 continue;
32:             add $\{g_1, g_2, g_3\}$ to the array of symmetry-preserving superlattices;
33: **return** $\{\{g_1, g_2, g_3\}\}$;

Figure 3. Algorithm for fast enumeration of symmetry preserving superlattices for systems other than triclinic.

### 2.3.3 Enumeration Algorithm for the Triclinic Crystal System

The triclinic system doesn't benefit from the above algorithm since all its superlattices preserve the point symmetry operations of the primitive lattice, namely the identity operation and sometimes the inversion operation. For triclinic systems we accelerate the search for superlattices for which $r_{lattice} \geq r_{min}$ by again considering one dimension at a time. For each factor set, if $H_{11}|\mathbf{a}_1| < r_{min}$, the shortest distance between lattice points must be less than $r_{min}$ and the factor set is not considered. Similarly, if the two dimensional lattice spanned by $H_{11}\mathbf{a}_1$ and $H_{21}\mathbf{a}_1 + H_{22}\mathbf{a}_2$ has $r_{lattice} < r_{min}$, we do not iterate over possible values of $H_{31}$ and $H_{32}$ as we already know the lattices will not satisfy the required constraint. The procedures are summarized as a pseudocode in Figure 4. The input lattice can be of any dimension up to three. We note that a similar approach can be used to accelerate any scheme based on iterating over lattices in HNF, such as the one developed by Morgan et al..[32]

**Algorithm 2** Fast Enumeration of Symmetry-Preserving Superlattices for Triclinic Structures

**Input:**
- $N_T$ - Size of the superlattice. Also the total number of $k$-points in the grid.
- $r_{min}$ - Minimum periodic distance specified by users.
- $\{\mathbf{a_1}, \mathbf{a_2}, \mathbf{a_3}\}$ - Primitive lattice vectors.

**Output:**
- $\{\{\mathbf{g_1}, \mathbf{g_2}, \mathbf{g_3}\}\}$ - Array of symmetry-preserving superlattices with $r_{lattice} \geq r_{min}$.

**Initialization:**
- $\mathbf{H}$ - A $3 \times 3$ zero matrix.
- $\{\{\mathbf{g_1}, \mathbf{g_2}, \mathbf{g_3}\}\}$ - Empty array of symmetry-preserving superlattices.

1: factorSets[ ][ ] ← sets of integral factorizations of $N_T$ into 3 numbers;
2: **for** $\{N_1, N_2, N_3\}$ in factorSets[ ][ ] **do**
3:     $\mathbf{H} \leftarrow$ put $N_1, N_2, N_3$ on diagonal positions;
4:     $\mathbf{g_1} = H_{11} \cdot \mathbf{a_1}$;
5:     **if** $\|\mathbf{g_1}\| < r_{min}$ **then**
6:         continue;
7:     **for** $H_{21} = 0$ to $N_1 - 1$ **do**
8:         $\mathbf{g_2} = H_{21} \cdot \mathbf{a_1} + H_{22} \cdot \mathbf{a_2}$;
9:         **if** $getMinDistance(\mathbf{g_1}, \mathbf{g_2}) < r_{min}$ **then**
10:            continue;
11:         **for** $(H_{31}, H_{32}) = (0, 0)$ to $(N_1 - 1, N_2 - 1)$ **do**
12:            $\mathbf{g_3} = H_{31} \cdot \mathbf{a_1} + H_{32} \cdot \mathbf{a_2} + H_{33} \cdot \mathbf{a_3}$;
13:            **if** $getMinDistance(\mathbf{g_1}, \mathbf{g_2}, \mathbf{g_3}) < r_{min}$ **then**
14:                continue;
15:            add $\{\mathbf{g_1}, \mathbf{g_2}, \mathbf{g_3}\}$ to the array of superlattices;
16: **return** the array of superlattices;

Figure 4. Algorithm for enumerating symmetry-preserving superlattices for triclinic system, accelerated by enforcing $r_{lattice} \geq r_{min}$ at each dimension.

## 2.4 Evaluating Shift Vectors

*K*-point grids can be generated for each symmetry-preserving lattice using equation (7), where the matrix $\mathbf{H}$ can be used for $\mathbf{M}$. The only remaining unknown is the shift vector $\mathbf{s}$. When the shift vector has zero length, the *k*-point grid is called a Γ-centered grid, as it must contain the Γ point in reciprocal space as a grid point. Often the use of shift vectors with non-zero length results in more efficient grids, in part because avoiding the highly-symmetric Γ point allows for greater use of symmetry to reduce the number of symmetrically irreducible *k*-points.

For a shift to be guaranteed to result in a symmetry-preserving lattice, it must shift the origin to a point that has the full point group symmetry of the origin. For all symmorphic space groups, the only such points are located at linear combinations of full- or half-multiples of the primitive lattice vectors. Thus, we consider only the eight such unique combination of *k*-point grid generating vectors, $\{\mathbf{d}_1, \mathbf{d}_2, \mathbf{d}_3\}$, as candidate shift vectors (Table 1). In some cases (e.g. hexagonal systems), some of the shift vectors in Table 1 will not result in a symmetry-preserving grid. We identify and reject these when determining the number of irreducible *k*-points. As this occurs as soon as the first point that breaks symmetry is encountered, it comes with relatively little computational cost.

**2.5 Algorithm for Fast Calculation of Symmetrically Irreducible K-points and K-point Weights**

We select the optimal lattice based on the values of $N_i$, $r_{lattice}$, and $N_T$. The value of $r_{lattice}$ can be easily obtained from the superlattice vectors by Minkowski reduction, and $N_T$ equals the absolute value of the determinant of the transformation matrix $\mathbf{M}$. However, calculating $N_i$ for a *k*-point grid is a relatively expensive operation. An intuitive approach is to apply all the point symmetry operations to each *k*-point, $\mathbf{k}_i$, and compare the resulting coordinates with all the other *k*-points. If one of the transformed *k*-points, $\mathbf{k}_i'$, is translationally equivalent to one of the other *k*-points, $\mathbf{k}_j$, then the *k*-points $\mathbf{k}_i$ and $\mathbf{k}_j$ are symmetrically equivalent. However, this algorithm scales as $O(N_T^2)$, where $N_T$ is the number of total *k*-points of a grid. As this operation is applied to each of the *k*-point grids found by the algorithm in section 2.3, this intuitive but costly approach could easily become the major overhead of any *k*-point generation scheme.

We solve this complication by first recognizing that a unit cell in reciprocal space is a supercell of a regular *k*-point lattice, where the two lattices are related by equation (6). To avoid confusion with the Hermite normal form of $\mathbf{M}$, which we have labelled $\mathbf{H}$, we will refer to the Hermite normal form of the transformation matrix in reciprocal space, $\mathbf{M}^T$, as $\mathbf{J}$ (in general, $\mathbf{J} \neq \mathbf{H}^T$). The key to our approach is the recognition that it is possible to tessellate all of reciprocal space with supercells of size $J_{11} \times J_{22} \times J_{33}$ arranged periodically on the superlattice, where $J_{11}$, $J_{22}$, and $J_{33}$ are the diagonal elements of $\mathbf{J}$ and each lattice point is a corner of the supercell. This is illustrated in two dimensions in Figure 5, but the same concept extends to any number of dimensions. The off-diagonal elements of $\mathbf{J}$ serve to shift each layer of supercells relative to the previous layer, so that the tessellation resembles stacked bricks. Within each of these supercells, the coordinates of a $k$-point can be expressed as:

$$\mathbf{r} + \left( [k_1, k_2, k_3] + \mathbf{s} \right) \cdot [\mathbf{d}_1, \mathbf{d}_2, \mathbf{d}_3]^T \qquad (12)$$

where $\mathbf{r}$ is a lattice point on the reciprocal space lattice (blue dots in Figure 5), $\mathbf{d}_1$, $\mathbf{d}_2$, and $\mathbf{d}_3$ are generating lattice vectors of the k-point lattice (also reciprocal primitive lattice vectors), $k_1$ is an integer from 0 to $J_{11} - 1$, $k_2$ is an integer from 0 to $J_{22} - 1$, and $k_3$ is an integer from 0 to $J_{33} - 1$. The coordinates of the k-point can then be easily transformed into any basis (such as that of the primitive lattice in reciprocal space) using linear operations. We have shared this approach for iterating over k-points with the Hart group for their work with generalized k-point grids.[31] Values for $k_1$, $k_2$, and $k_3$ can be quickly calculated for any k-point using integer arithmetic, as discussed below and shown in lines 15 and 16 of Figure S6 of Supplementary Information.

Given the enumeration of $k$-points using equation (12), we identify irreducible $k$-points in a way similar to that described by Hart et al..[31] We assign a unique index to each $k$-point in the Brillouin zone or, equivalently, to each $k$-point in any unit cell of the reciprocal lattice, by

$$index = 1 + k_1 + J_{11}k_2 + J_{11}J_{22}k_3. \qquad (13)$$

The values of the index range from 1 to $N_T$, and translationally equivalent k-points share the same index. Linear scaling is achieved because the index for any given k-point can be calculated in constant time, as can the sublattice of k-points that have a given index. Then iteration of all k-points in a unit cell in reciprocal space, equivalent to all k-points in the Brillouin zone, is accomplished by looping over values of $k_1$, $k_2$, and $k_3$ in equation (13).

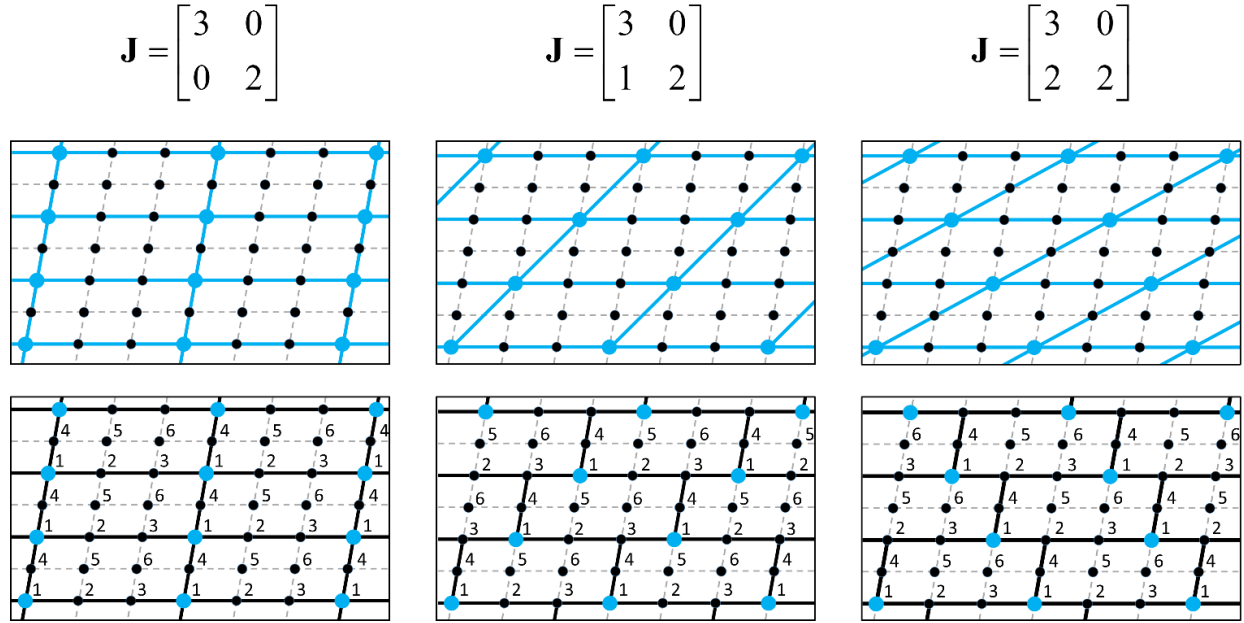

Figure 5. Two-dimensional illustrations of the concepts used for k-point enumeration and index generation. The top row provides the three possible matrices in Hermite normal form for the set of factors (3,2). The middle row shows the three Bravais superlattices corresponding to these matrices, assuming that the generating vectors for the k-point grid, $\mathbf{d}_1$ and $\mathbf{d}_2$, are aligned with the dashed gray lines. The bottom row shows how space can be tessellated by unit cells that are $3 \times 2$ supercells of the generating lattice vectors, with k-point indices marked within each cell.

To count the number of distinct k-points, we iterate over all translationally distinct k-points as described above and apply all symmetry operations to each k-point. If an operation does not transform the *k*-point to another *k*-point, the grid is not symmetry-preserving and is rejected (this can sometimes happen if a shift of the $\Gamma$ point breaks symmetry). If the index of any symmetrically equivalent *k*-point is less than that of the current *k*-point, then we have already seen a symmetrically equivalent *k*-point, so the counter for the number of irreducible *k*-points is not incremented. If there is no symmetrically equivalent *k*-point with an index lower than that of the current *k*-point, then the current *k*-point is the first we have seen in its orbit, so the counter for the number of irreducible *k*-points is incremented. A simple variation of this algorithm is used to calculate *k*-point weights by, for each *k*-point, determining the orbit of symmetrically equivalent points and then incrementing the weight of the *k*-point that has the lowest index in that orbit. Figure S6 in supporting information provides the pseudocode of this algorithm. The final, returned arrays contain coordinates and weights for all *k*-points. The symmetrically non-distinct points, however, have weights of zero. This fact is used to identify the subset of irreducible points.

### 3. KpLib: A Lightweight, Open-source C++ Library

To facilitate the integration of the generalized Monkhorst-Pack *k*-point grids in simulation packages, we implemented the presented algorithms in a lightweight library, kpLib. It is written in C++ to make interfacing easier for as many programming languages as possible. A python module, kpGen, is also provided as a wrapper of the C++ library. The source code of kpLib only contains 1122 lines, and the API uses elementary data structures as argument types, which should be available in most programming languages and facilitate the construction of wrapping functions. We have written a demonstration application, integrated with *spglib* [38], to show how to work with the API. The library is open sourced and a documentation of the API is provided on the homepage of its public repository (https://gitlab.com/muellergroup/kplib). We note that packages that plan to integrate kpLib should ensure that the set of symmetry operations used to generate the *k*-point grid are used consistently in the rest of the code.

## 4. Benchmarks

Here we present a series of benchmarks to demonstrate the speed at which our algorithm generates *k*-point grids and the efficiency of the generated grids, including a comparison to the grids generated using GRkgridgen.[31, 32] All benchmarks were performed on the 102 structures randomly selected from the Inorganic Crystal Structure Database (ICSD) used in our previous work [2, 30]. Version 2019.09.17 for kpLib was used for all benchmarks.

### 4.1 Grid Generation Speed

We have benchmarked the speed at which kpLib generates both Γ-centered grids and grids with automatically selected shift vectors (called "auto grids" in the following text). To accelerate searches for large grids, we use an approach in which a search for small grids is performed, and then the densities of the small grids are increased in every dimension by a constant scale factor. This use of the scale factor was first introduced in section II.D of our previous work,[2] and it is also adopted in the dynamic generation approach (for a detailed discussion, see section 1 of Supplementary Information). We have benchmarked grid generation speed on 102 randomly selected structures using a single core on Intel Xeon E5660 processors with a 2.80 GHz base frequency and a 48 GB RAM, with and without the use of the scale factor. Grid sizes are specified by $r_{min}$, instead of $N_{min}$, as the former is physically more meaningful,[2, 4] and thus we believe it is the most likely method to be used. A benchmark using $N_{min}$ to compare the speed of the dynamic generation approach and the database look-up approach is given in section 7 of Supplementary Information.

Average computation time for both Γ-centered grids and auto grids are shown in Figure 6. The speed at which kpLib generates Γ-centered and auto grids is very similar. When $r_{min}$ is 50 angstroms, which is sufficient for converging most calculations within 1 meV / atom,[2] both types of grids are generated in less than 0.2 seconds on average. For large grids, using the scale factor increases generation speed, at a slight cost of grid quality (Figure 8). When $r_{min}$ is 100 angstroms, it takes only about 1 second to find the optimal grids using the scale factor, while the exhaustive search with scale factor switched off finishes in about 4.6 seconds. The smallest value of $r_{min}$ at which the scale factor starts to have an effect is 55 angstroms, but not all 102 structures use the scale factor at 55 angstroms and 69 out of the 102 structures do not use the scale factor even at 100 angstroms.

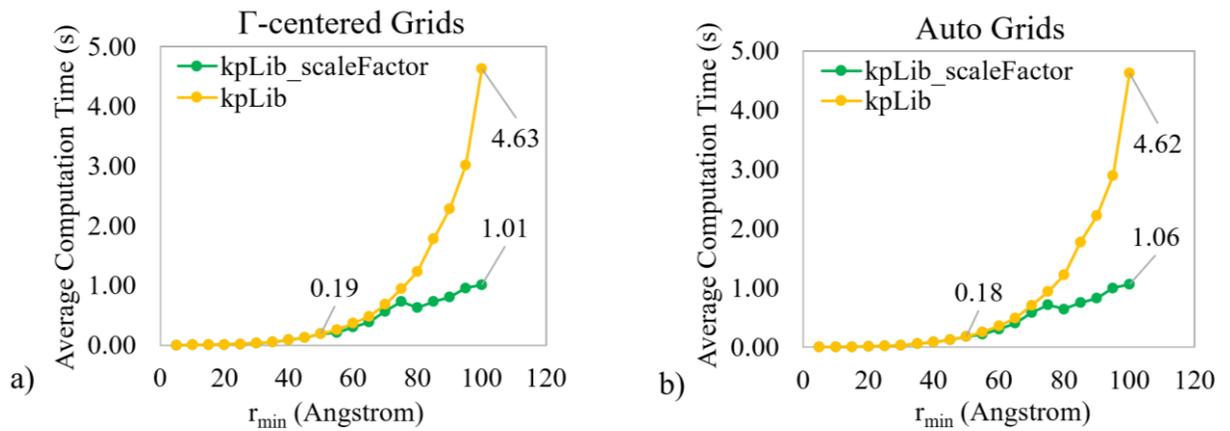

Figure 6. Average computation time of dynamic generation using kpLib with and without scale factors at various values of $r_{min}$ for a) Γ-centered grids and b) auto grids. The computation times at $r_{min}$ = 100 angstroms are labeled on the graphs.

The dynamic generation approach used by kpLib is more lightweight than the database approach used by the *K*-Point Grid Generator, which includes a 7.3 MB database containing 428,632 pre-generated grids. However the database lookup method (section 3 of Supplementary Information) is generally faster (Figure 7). Database searching is much faster than dynamic grid generation for Γ-centered grids over a wide range of densities. The difference between the two approaches is smaller when shifted grids are included, but the database is still two times as fast at the largest $r_{min}$. This difference in relative performance for shifted grids can be attributed to the fact that dynamic grid generation loops over $N_T$, and the database search loops over $N_i$. When searching for shifted grids rather than only Γ-centered grids, the upper bound for the loop over $N_T$ is more rapidly reduced due to the larger number of candidate grids (Figure 1), whereas the upper bound for the loop over $N_i$ is not (Figure S3 of the Supplementary Information).

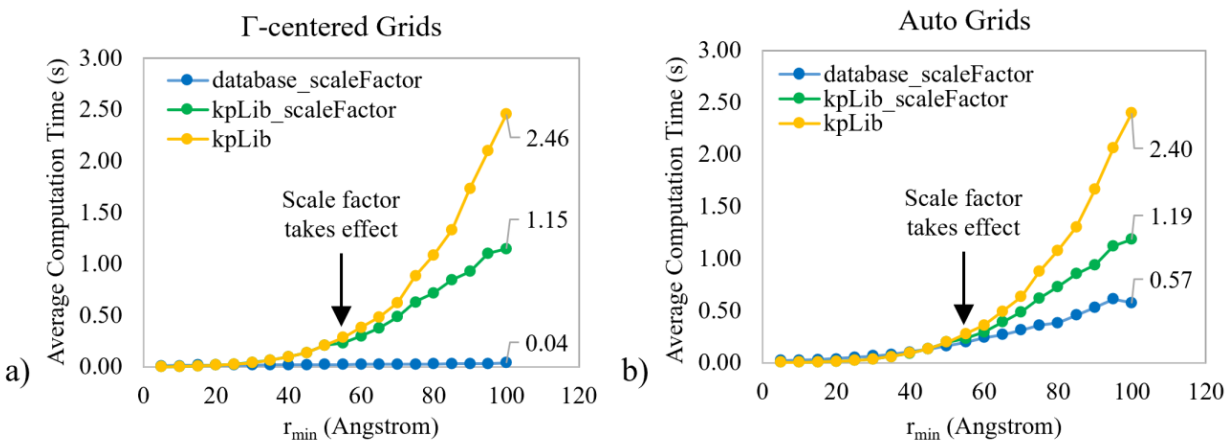

Figure 7. Comparison of computation time between database lookup method used by the *K*-Point Grid Generator and the dynamic generation approach used by kpLib. This benchmark did not include monoclinic and triclinic structures, as both the *K*-Point Grid Generator and kpLib use dynamic grid generation for them.

**4.2 Grid Quality Comparison between KpLib and GRkgridgen**

We compared our dynamic grid generation method with GRkgridgen, another software package which can generate generalized Monkhorst Pack grids.[32] As the options for grid generation differ between the two packages, we used the following settings to make a fair comparison: both applications were instructed to select the grid with minimal $N_i$ (a natural measure of the efficiency of a grid that meets user-provided constraints), and the required *k*-point density was specified by providing a value for $N_{min}$ (defined as MINTOTALKPOINTS in kpLib and NKPTS in GRkgridgen). In the version we tested, 0.7.5, GRkgridgen doesn't guarantee that the real-space superlattices corresponding to the returned grids satisfy $r_{lattice} \geq r_{min}$, but it does take $r_{lattice}$ into account when generating grids based on $N_{min}$. As kpLib only accounts for $r_{lattice}$ if $r_{min}$ is provided by the user, to ensure a fair comparison we have constrained the grids generated by kpLib, to have $r_{lattice}$ which is at least as large as that of the grid generated by GRkgridgen at the same $N_{min}$ and for the same structure. The same 102 structures were used and both $\Gamma$-centered grids and auto grids were compared. For kpLib without a scale factor, $N_{min}$ values ranged from 1 to 5623, while for kpLib using scale factor, the range is increased to 15,848 to better demonstrate the effect of scale factor for large grids.

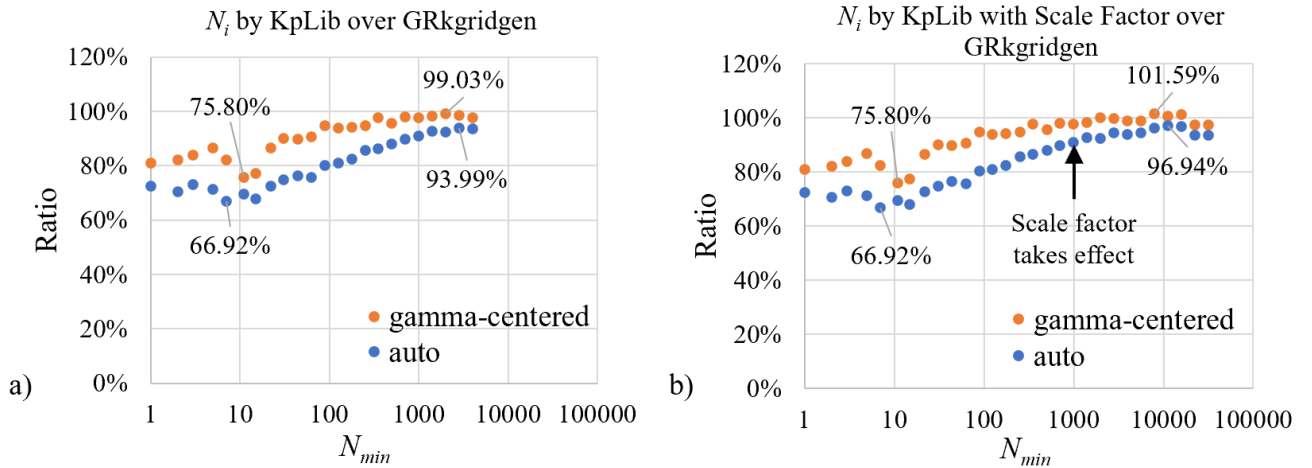

Figure 8. Ratios of average number of symmetrically irreducible *k*-points from the dynamic search by a) kpLib, b) kpLib with the scale factor, to grids generated using GRkgridgen, for both $\Gamma$-

centered grids and auto grids. Both the maximal and minimal ratios are labeled for both types of grids. Part b) has a larger range of $N_{min}$ (from 1 to 15,848), to better demonstrate the effect of the scale factor on grid quality.

We use the number of irreducible *k*-points in the generated grid as a metric of grid efficiency, as the computational cost of most calculations that use *k*-points scales linearly with the number of irreducible *k*-points. The scale factor makes little difference in the number of irreducible *k*-points for grids generated below $N_{min} = 5623$ (Figure 8). For auto grids at all values of $N_{min}$, including those generated using the scale factor, grids from kpLib consistently have fewer irreducible *k*-point than the grids from GRkgridgen on average. The same is true for $\Gamma$-centered grids generated without using the scale factor, although for very dense grids when the scale factor is used GRkgridgen may return grids that are 1-2% more efficient on average. The difference between kpLib and GRkgridgen is much larger for auto grids than $\Gamma$-centered grids, and it is larger for small $N_T$ than large ones. We note that the gain in performance for relatively small values of $N_T$ can be particularly beneficial as calculations with such small grids often have large supercells and are thus computationally demanding. For auto grids, which we expect to be the most commonly used mode, the expected increase in calculation speed using the grids generated by kpLib ranges from 3% to 37%.

## 5. Conclusion

The widespread use of generalized Monkhorst-Pack *k*-point grids has been limited by the lack of algorithms and tools for rapidly generating highly efficient grids. By effectively reducing the problem of generating optimal 3-dimenstional generalized Monkhorst-Pack *k*-point grids to that of enumerating over 2-dimensional lattices, along with several other algorithmic innovations, we have demonstrated that is possible to very rapidly identify optimal generalized Monkhorst-Pack *k*-point grids for a given material, given user constraints on the spacing of the real-space grid points and/or the minimum total required *k*-points. For commonly-used grid densities, the grids generated by the algorithms presented in this paper are on average significantly more efficient than those generated using previously developed algorithms. Given the demonstrated benefits of using generalized Monkhorst-Pack *k*-point grids[2-4, 24], we conservatively estimate that widespread adoption of these algorithms could save computational materials researchers more than a hundred million CPU hours, worth millions of US dollars, each year. To facilitate this widespread use, we have implemented our algorithms for grid generation in kpLib, a lightweight open source library with only 1122 lines of code for integration with third-party software algorithms, and we have developed a standalone open-source tool, the *K*-Point Grid Generator, for rapidly generating generalized Monkhorst-Pack grids.

**Corresponding Author**

Tim Mueller – Department of Materials Science and Engineering, Johns Hopkins University, Baltimore, MD 21218, United States. E-mail: tmueller@jhu.edu

**Data Availability**

The source code of kpLib and *K*-Point Grid Generator are freely available online at https://gitlab.com/muellergroup/kplib and https://gitlab.com/muellergroup/k-pointGridGenerator. The raw data required to reproduce these findings are available to download from https://gitlab.com/muellergroup/kplib/-/blob/master/doc/paper_data/Raw_Data.xlsx.

## ACKNOWLEDGMENTS

Y.W., P.W., A.B., and T.M. thank the National Science Foundation for the financial support under Award No. DMR-1352373, the Homewood High-Performance Cluster (HHPC) and Maryland Advanced Research Computing Center (MARCC) for providing computational resources, and Prof. Gus Hart for helpful discussions. S.D. thanks the U.S. Department of Energy, Office of Science, Office of Basic Energy Sciences, Materials Sciences and Engineering Division for financial support under Contract no. DE-AC02-05-CH11231: Materials Project program KC23MP.

## REFERENCES

1. Monkhorst, H. J.; Pack, J. D., Special Points for Brillouin-Zone Integrations. *Physical Review B* **1976,** *13* (12), 5188-5192.
2. Wisesa, P.; McGill, K. A.; Mueller, T., Efficient generation of generalized Monkhorst-Pack grids through the use of informatics. *Physical Review B* **2016,** *93* (15).
3. Morgan, W. S.; Jorgensen, J. J.; Hess, B. C.; Hart, G. L. W., Efficiency of Generalized Regular k-point grids. *Comp Mater Sci* **2018,** *153*, 424-430.
4. Choudhary, K.; Tavazza, F., Convergence and machine learning predictions of Monkhorst-Pack k-points and plane-wave cut-off in high-throughput DFT calculations. *Comp Mater Sci* **2019,** *161*, 300-308.
5. Amazon Web Services Amazon EC2 Pricing. https://aws.amazon.com/ec2/pricing/on-demand/ (accessed June 19 2020).
6. Froyen, S., Brillouin-zone integration by Fourier quadrature: Special points for superlattice and supercell calculations. *Phys Rev B* **1989,** *39* (5), 3168-3172.
7. Moreno, J.; Soler, J. M., Optimal Meshes for Integrals in Real-Space and Reciprocal-Space Unit Cells. *Physical Review B* **1992,** *45* (24), 13891-13898.
8. Mundet, B.; Hartman, S. T. T.; Guzman, R.; Idrobo, J. C. C.; Obradors, X.; Puig, T.; Mishra, R.; Gazquez, J., Local strain-driven migration of oxygen vacancies to apical sites in YBa2Cu3O7-x. *Nanoscale* **2020,** *12* (10), 5922-5931.
9. Chowdhury, T.; Kim, J.; Sadler, E. C.; Li, C.; Lee, S. W.; Jo, K.; Xu, W.; Gracias, D. H.; Drichko, N. V.; Jariwala, D.; Brintlinger, T. H.; Mueller, T.; Park, H.-G.; Kempa, T. J., Substrate-directed synthesis of MoS2 nanocrystals with tunable dimensionality and optical properties. *Nature Nanotechnology* **2020,** *15* (1), 29-34.
10. Wisesa, P.; Li, C.; Wang, C.; Mueller, T., Materials with the CrVO4 structure type as candidate superprotonic conductors. *RSC Advances* **2019,** *9* (55), 31999-32009.
11. Williams, L.; Kioupakis, E., BAlGaN alloys nearly lattice-matched to AlN for efficient UV LEDs. *Appl Phys Lett* **2019,** *115* (23), 231103.
12. Wang, Y.; Cao, L.; Libretto, N. J.; Li, X.; Li, C.; Wan, Y.; He, C.; Lee, J.; Gregg, J.; Zong, H.; Su, D.; Miller, J. T.; Mueller, T.; Wang, C., Ensemble Effect in Bimetallic Electrocatalysts for CO2 Reduction. *J Am Chem Soc* **2019,** *141* (42), 16635-16642.

13. Rosenbrock, C. W.; Gubaev, K.; Shapeev, A. V.; Pártay, L. B.; Bernstein, N.; Hart, G. L., Machine-learned Interatomic Potentials for Alloys and Alloy Phase Diagrams. *arXiv preprint arXiv:1906.07816* **2019**.
14. Nyshadham, C.; Rupp, M.; Bekker, B.; Shapeev, A. V.; Mueller, T.; Rosenbrock, C. W.; Csányi, G.; Wingate, D. W.; Hart, G. L., Machine-learned multi-system surrogate models for materials prediction. *Npj Comput Mater* **2019,** *5* (1), 51.
15. Li, C.; Gao, H.; Wan, W.; Mueller, T., Mechanisms for hydrogen evolution on transition metal phosphide catalysts and a comparison to Pt(111). *Phys Chem Chem Phys* **2019,** *21* (44), 24489-24498.
16. Kratzer, P.; Neugebauer, J., The basics of electronic structure theory for periodic systems. *Frontiers in chemistry* **2019,** *7*.
17. Hernandez, A.; Balasubramanian, A.; Yuan, F.; Mason, S. A. M.; Mueller, T., Fast, accurate, and transferable many-body interatomic potentials by symbolic regression. *Npj Comput Mater* **2019,** *5* (1), 112.
18. Greenman, K.; Williams, L.; Kioupakis, E., Lattice-constant and band-gap tuning in wurtzite and zincblende BInGaN alloys. *Journal of Applied Physics* **2019,** *126* (5), 055702.
19. Cao, L.; Zhao, Z.; Liu, Z.; Gao, W.; Dai, S.; Gha, J.; Xue, W.; Sun, H.; Duan, X.; Pan, X.; Mueller, T.; Huang, Y., Differential Surface Elemental Distribution Leads to Significantly Enhanced Stability of PtNi-Based ORR Catalysts. *Matter* **2019,** *1* (6), 1567-1580.
20. Cao, L.; Niu, L.; Mueller, T., Computationally generated maps of surface structures and catalytic activities for alloy phase diagrams. *Proceedings of the National Academy of Sciences of the United States of America* **2019,** *116* (44), 22044-22051.
21. Liu, Y.; Zhang, H.; Behara, P. K.; Wang, X.; Zhu, D.; Ding, S.; Ganesh, S. P.; Dupuis, M.; Wu, G.; Swihart, M. T., Synthesis and Anisotropic Electrocatalytic Activity of Covellite Nanoplatelets with Fixed Thickness and Tunable Diameter. *ACS applied materials & interfaces* **2018,** *10* (49), 42417-42426.
22. Li, C. Y.; Raciti, D.; Pu, T. C.; Cao, L.; He, C.; Wang, C.; Mueller, T., Improved Prediction of Nanoalloy Structures by the Explicit Inclusion of Adsorbates in Cluster Expansions. *J Phys Chem C* **2018,** *122* (31), 18040-18047.
23. Ding, Y.; Wang, Y., Tunable Electronic Structures of Hydrogenated Zigzag and Armchair Dumbbell Silicene Nanosheets: A Computational Study. *The Journal of Physical Chemistry C* **2018,** *122* (40), 23208-23216.
24. Wolloch, M.; Suess, D.; Mohn, P., Influence of antisite defects and stacking faults on the magnetocrystalline anisotropy of FePt. *Physical Review B* **2017,** *96* (10), 104408.
25. Williams, L.; Kioupakis, E., BInGaN alloys nearly lattice-matched to GaN for high-power high-efficiency visible LEDs. *Appl Phys Lett* **2017,** *111* (21), 211107.
26. Raciti, D.; Cao, L.; Liv, K. J. T.; Rottmann, P. F.; Tang, X.; Li, C. Y.; Hicks, Z.; Bowen, K. H.; Hemker, K. J.; Mueller, T.; Wang, C., Low-Overpotential Electroreduction of Carbon Monoxide Using Copper Nanowires. *Acs Catalysis* **2017,** *7* (7), 4467-4472.
27. Cao, L.; Raciti, D.; Li, C. Y.; Livi, K. J. T.; Rottmann, P. F.; Hemker, K. J.; Mueller, T.; Wang, C., Mechanistic Insights for Low-Overpotential Electroreduction of $CO_2$ to CO on Copper Nanowires. *Acs Catalysis* **2017,** *7* (12), 8578-8587.
28. Vienna Ab initio Simulation Package KPOINTS in VASP Wiki. https://www.vasp.at/wiki/index.php/KPOINTS.

29. Li, C.; Nilson, T.; Cao, L.; Mueller, T., Predicting activation energies for vacancy-mediated diffusion in alloys using a transition-state cluster expansion. *arXiv preprint arXiv:2009.12474* **2020**.
30. Inorganic Crystal Structure Database. Fiz Karlsruhe: 2016.
31. Hart, G. L. W.; Jorgensen, J. J.; Morgan, W. S.; Forcade, R. W., A robust algorithm for k-point grid generation and symmetry reduction. *Journal of Physics Communications* **2019**, *3* (6), 065009.
32. Morgan, W. S.; Christensen, J. E.; Hamilton, P. K.; Jorgensen, J. J.; Campbell, B. J.; Hart, G. L. W.; Forcade, R. W., Generalized regular k-point grid generation on the fly. *Comp Mater Sci* **2020**, *173*.
33. Ashcroft, N. W.; Mermin, N. D., *Solid State Physics*. Brooks/Cole: Belmont, USA, 1976; p 132-143.
34. Tinkham, M., *Group Theory and Quantum Mechanics*. McGraw-Hill Book Company: New York, 1964; p 279-281.
35. Giacovazzo, C.; Monaco, H. L.; Artioli, G.; Viterbo, D.; Milanesio, M.; Ferraris, G.; Gilli, G.; Gilli, P.; Zanotti, G.; Catti, M., *Fundamentals of Crystallography*. third ed.; Oxford University Press: Oxford, 2002; p 842.
36. Mueller, T. Computational studies of hydrogen storage materials and the development of related methods. Massachusetts Institute of Technology, Boston, Massachusetts, 2007.
37. Hart, G. L. W.; Forcade, R. W., Algorithm for generating derivative structures. *Physical Review B* **2008**, *77* (22).
38. Togo, A.; Tanaka, I., Spglib: a software library for crystal symmetry search. *arXiv preprint arXiv:1808.01590* **2018**.

# Rapid Generation of Optimal Generalized Monkhorst-Pack Grids

# Supplementary Information

Yunzhe Wang[1], Pandu Wisesa[1], Adarsh Balasubramanian[1], Shyam Dwaraknath[2] and Tim Mueller[1,*]

[1] Department of Materials Science and Engineering, Johns Hopkins University, Baltimore, Maryland 21218, USA

[2] Lawrence Berkeley National Laboratory, Berkeley, California, 94720, USA

E-mail: tmueller@jhu.edu

The supplementary information is organized in the following way: firstly, the dynamic generation method with scale factor is discussed; secondly, the database search method is explained, which uses a pre-generated database of optimal grids up to fixed sizes to overcome the hurdle of expansiveness of exhaustive search; then, a few algorithms, not essential but useful to the generation of generalized Monkhorst-Pack grids, are introduced; next, additional implementations of algorithms and methods presented in this paper is reviewed; lastly, a comparison of the speed performance of the dynamic generation method and the database lookup method is shown.

## 1. Using a Scale Factor for Dense Grids

For grids with a large number of total *k*-points, a fully dynamic search is computationally expensive, especially for the triclinic crystal system. To compromise between the speed and grid quality, we introduced scale factor in section II. D of our previous work [1]. The basic idea behind this approach is that rather going through the computationally expensive process of trying to find the optimal grid for some large value of $N_T$, we instead do a much faster search for a grid with $N_T/n^3$ total *k*-points, where the scale factor $n$ is a positive integer[1]. The periodic lattice vectors for the real-space superlattice for this grid are then multiplied by $n$ to construct a grid with $N_T$ total *k*-points. This approach is necessary when generating grids using the database due to the finite size of the database, and details of how it is implemented for database-generated grids is shown in following sections.

The scale factor takes effect in dynamic grid generation when the maximum search depth has been reached and no qualifying grid has been found. In this case the scale factor, which is initialized to a value of 1, is incremented and a new iteration is started with a lower bound of grid sizes calculated by

$$N_{lower} = \max\left(\frac{N_{min}}{n^3}, \left\lfloor \frac{\sqrt{2}}{2}\left(\frac{r_{min}}{n}\right)^3 \Big/ V_p \right\rfloor\right). \tag{1}$$

The upper bound is reset to the maximum search depth. When a grid satisfying all constraints is found, the upper bound is updated by:

$$N_{upper} = \frac{N_i \times N_{sym}}{n^3} \tag{2}$$

The candidate grids are all scaled back by the scale factor before evaluating the values of $R_{lattice}$ and $N_i$ for assessing grid quality.

The default maximum search depths are 729 ($9\times9\times9$), 1729 ($12\times12\times12$), 46656 ($36\times36\times36$) and 5832 ($18\times18\times18$) for triclinic, monoclinic, cubic and the other four crystal systems. Users can change these values in the code if they desire different limits. The search stops if the scale factor becomes larger than 3. Therefore, if the scale factor is used, the maximum size of a grid that can be returned is 27 times the maximum depth. If best grid is still not found, a message is displayed to remind users that the request exceeds the current maximum search capability. Figure S1 illustrates the workflow of the dynamic search using scale factors.

---

[1] The symbols used in the supplementary information share the same definitions of the symbols appearing in the main text.

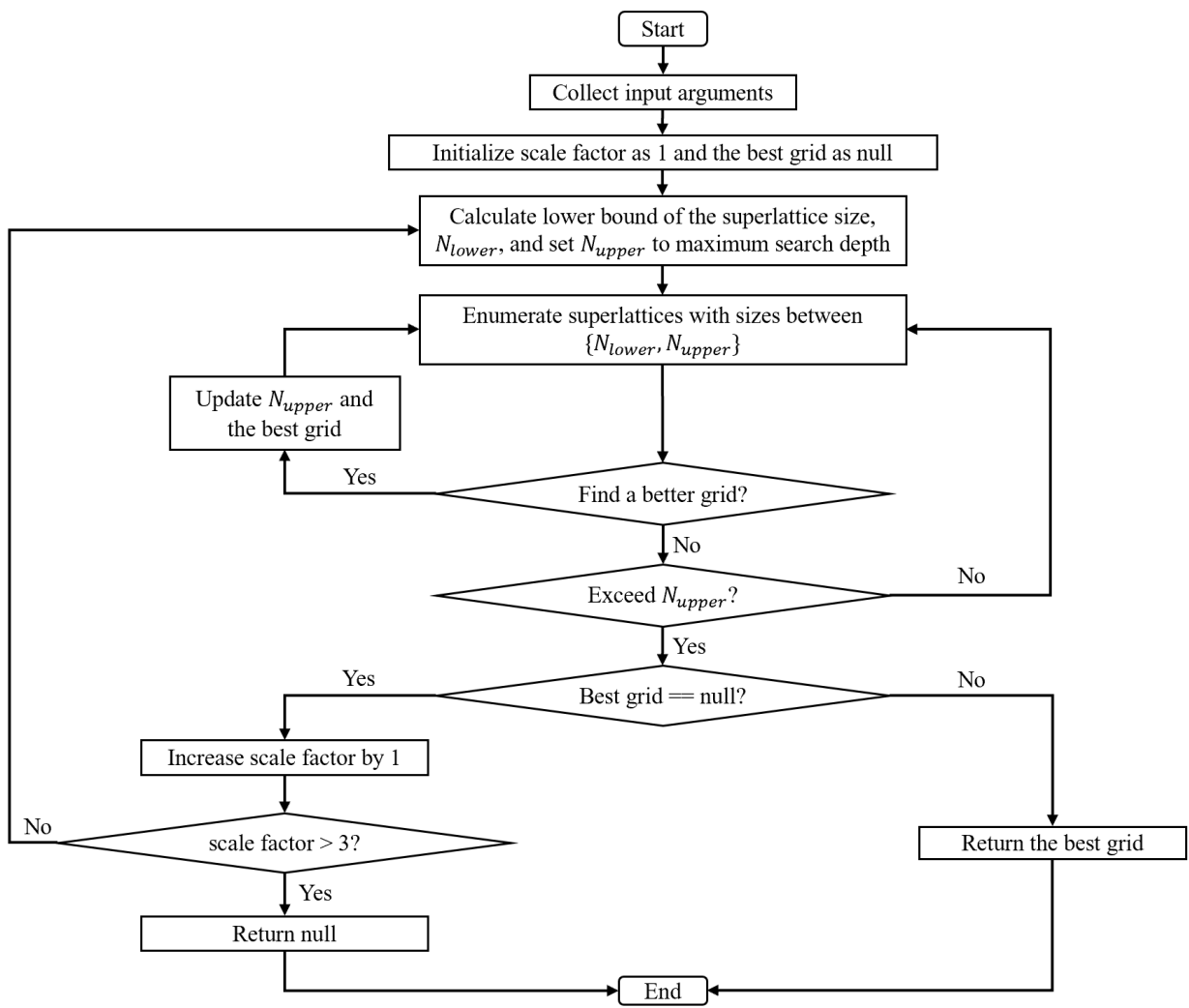

Figure S1. Workflow of the dynamic grid generation with scale factor activated.

## 2. Estimation of the Number of Possible Superlattices at a Given Size

Given a size, $N_T$, of a superlattice, and a factorization of it into three distinct positive integers $\{N_1, N_2, N_3\}$, the number of unique matrices in Hermite Normal Form (HNF) with the three numbers as the diagonal elements, would be

$$N_1^2 N_2 + N_2^2 N_1 + N_1^2 N_3 + N_3^2 N_1 + N_2^2 N_3 + N_3^2 N_2, \tag{3}$$

counting all permutations of the three numbers at the diagonal positions. The total number of possible superlattices can be calculated by considering all ways in which $N_T$ can be factored into three numbers.

## 3. An Overview of the Database Search Approach

Our previous work uses a database of pre-generated optimal grids to accelerate the search and make the generation of generalized Monkhorst-Pack grids feasible [1]. With the new and faster dynamic generation method, the database has been updated to include grids up to larger sizes. The database currently contains 428,632 pre-calculated symmetry-preserving grids, both shifted and Γ-centered, covering each of the 21 centrosymmetric symmorphic space groups other than triclinic and monoclinic ones. The number of grids has increased by 637%, compared with previous version of the database, as we have increased grid density and the number of shift vectors considered. The maximum size of stored *k*-point grids has increased from 1,728 (12×12×12) to 5,832 (18×18×18) for orthorhombic, tetragonal, trigonal and hexagonal systems. The maximum size for cubic systems has grown to 46,656 (36×36×36). The grids for each of these space groups are stored in 42 separate binary files (21 for shifted and 21 for Γ-centered grids). Figure S2 gives a schematic outline of the database organization. The database groups grids with the same $N_i$ and the same symmetry group in one array, and indexes arrays by $N_i$ to accelerate the search for grids with minimal $N_i$. Each grid in the database has fields for $N_T$, $N_i$, the generating matrix **H** of its corresponding superlattice, the shift vector, and a set of coefficients for fast estimation of $r_{lattice}$. A unique index is used to represent **H**, generated using the same mechanism as used for iterating over superlattices described in section 2.3. The matrix can be easily recovered from the index. Memory is saved by storing only an integer instead of an array of nine numbers.

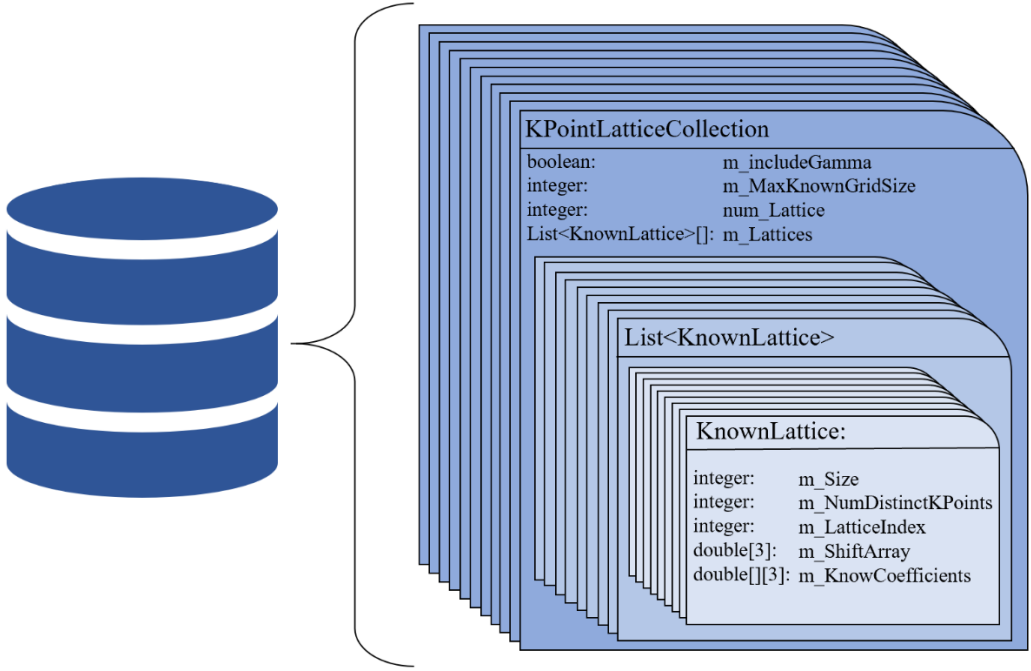

Figure S2. Schematic representation of the database organization. "m_IncludeGamma" specifies whether the grids in this file contain the $\Gamma$ point. "m_MaxKnownGridSize" indicates the maximum size of the grids. "num_Lattice" is the total number of k-point grids stored in this file. "m_Lattices" is an array of lists of grids. Grids with the same $N_i$ are stashed in the same list, and the lists are ordered by $N_i$. Each "KnownLattice" entry in the list represents a k-point grid. "m_Size" and "m_NumDistinctKPoints" represent the total number of k-points ($N_T$) and the number of symmetrically irreducible k-points ($N_i$). "m_LatticeIndex" is a unique index assigned to each superlattice for regenerating the transformation matrix in HNF, **H** (described in section 2.1 of supplementary information). "m_ShiftArray" stores the shift vector for the grid in coordinates of the reciprocal lattice for the conventional primitive cell. "m_KnownCoefficients" is an array of coefficients for quickly determining an upper bound for $r_{lattice}$.

In the following sections, we shall discuss in detail the procedures of the database searching method, and the algorithms for recovering a k-point grid from information stored in each entry.

### 3.1 Grid Generation by Searching the Database

Dynamic grid generation can be slow for requests with large $r_{min}$ and $N_{min}$. A pre-generated database accelerates grid generation by skipping the non-symmetry-preserving superlattices and the low quality superlattices (e.g. the ones with too many symmetrically irreducible k-points).

Figure S3 provides the workflow of the database search approach for generating the optimal generalised *k*-point grid. A procedure similar to that of the dynamic search is used. The difference is that the iteration of grids changes from explicitly constructing grids at each value of $N_T$ to a simple, constant-time lookup of entries in each array indexed by $N_i$. It starts by estimating the lower bound of the number of symmetrically irreducible *k*-points:

$$N_{lower} = m\_MinDistinctKPoints\left[\max\left(\frac{\sqrt{2}}{2}\cdot\left(\frac{r_{min}}{n}\right)^3\Big/V_p, \frac{N_{min}}{n^3}\right)\right]. \quad (4)$$

where $n$ is the scale factor and $m\_MinDistinctKPoints[]$ is an integral array created when loading a file from the database. The $N$-th element represents the minimum value of $N_i$ of all the grids stored in this file that have a size of $N$. The minimum $N$ that satisfies $r_{lattice} \geq r_{min}$ and $N \geq N_{min}$ is calculated by the $max()$ function. The first argument is the minimum size of a superlattice that could satisfy $r_{lattice} \geq r_{min}$, and the justification of the prefactor is similar to that of equation (10) of the main text. For Γ-centered grids of all lattices and shifted grids of non-cubic lattices, the minimum volume is that of a fcc unit cell with a distance of no more than $r_{min}$ between lattice points. For cubic systems, a fcc superlattice results in a bcc reciprocal lattice, and the only symmetry-preserving shift of the Γ point in a bcc lattices is $[0.5, 0.5, 0.5]$, which results in a Γ-centered grid. Therefore, the minimum volume of supercell for a shifted grid in a cubic system is that of the second closest-packed lattice, a body-centered-cubic (bcc) lattice [2]. $N_{lower}$ in this case is calculated by

$$N_{lower} = m\_MinDistinctKPoints\left[\max\left(\frac{4}{3\sqrt{3}}\cdot\left(\frac{r_{min}}{n}\right)^3\Big/V_p, \frac{N_{min}}{n^3}\right)\right]. \quad (5)$$

$N_{upper}$ is the maximum number of $N_i$ of grids in the database, which have the same symmorphic space group as the input structure. Grids are searched by starting with the list for which $N_i = N_{lower}$ and incrementally increasing the size of $N_i$ until $N_i = N_{upper}$. Once an optimal lattice for a given value of $N_i$ is found, there is no need to search for larger values of $N_i$. If the search cannot find a fulfilling grid before $N$ exceeds $N_{upper}$, the scale factor is increased by 1, the lower bound of the search is updated, and the search restarts from the new $N_{lower}$. If a grid satisfying all constraints is still not found when the scale factor exceeds 3, the search stops and a message is displayed to notify the user that the request exceeds the current maximum search capability. Since the database is generated by iterating grids up to at least the search depths mentioned at the beginning of this section, the maximum search capabilities are at least 27 times 46,656 ($36\times36\times36$) for cubic systems and 27 times 5,832 ($18\times18\times18$) for the other four crystal systems in the database. The alignment of search depths between database and the dynamic search with the scale factor ensures consistent results from both types of generation approaches.

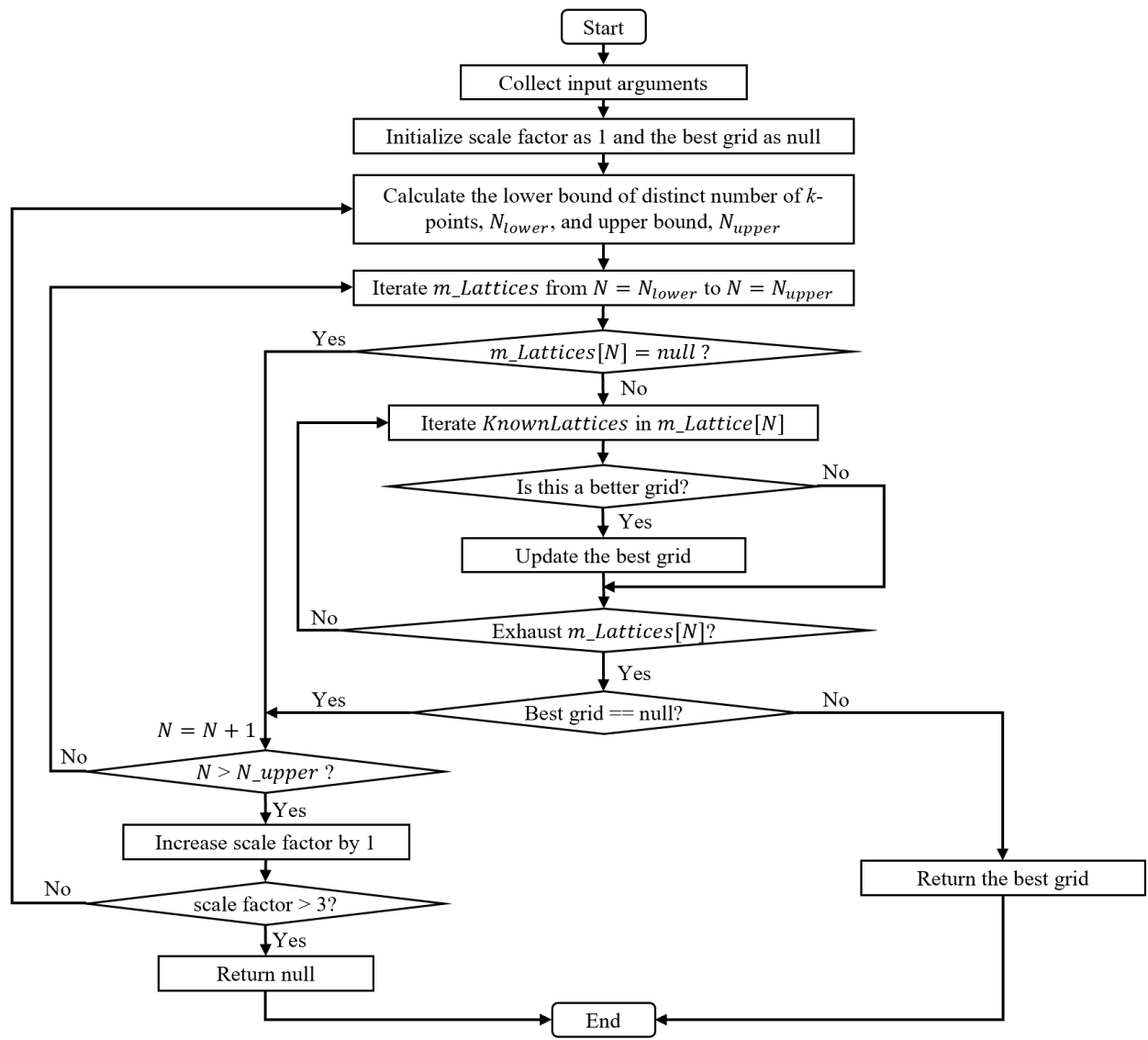

Figure S3. A diagram summarizes the grid generation workflow by the database-searching approach.

There are two situations in which the database may return slightly different grids than dynamic grid generation. The first is due to a difference in the way the scale factor is used for the database search and how it is used for dynamic grid generation. In dynamic grid generation, the optimized grid is chosen based on the value of $N_i$ for the final returned grid. For the database search, the value of $N_i$ for the scaled (smaller) grid is used, as the database is indexed by this value. In some cases this can result in the database returning a final grid that has a larger $N_i$ than the grid generated dynamically when the scale factor is used. The database may also return different results from dynamic grid generation when $N_{min}$ is the limiting factor in grid generation, as the database was generated using Pareto frontiers based on $r_{min}$, which we expect to be the more commonly-used constraint. In both situations any difference in efficiency between the grid returned by the database and the dynamically generated grid is typically small.

### 3.2 Storing and Retrieving K-point Grids from the Database

In this sector, we present algorithms for recovering grids from information stored in database entries: mapping each $\mathbf{H}$ to a unique index and rapidly estimating $r_{lattice}$ from sets of coefficients.

### 3.2.1 Assigning Superlattice Indices to $\mathbf{H}$ and Regenerating $\mathbf{H}$ from an Index

As shown by section 2.1 of the main text, each *k*-point grid uniquely corresponds to a real-space superlattice. Each superlattice can be uniquely represented by a transformation matrix $\mathbf{H}$, which is $\mathbf{M}$ in Hermite normal form [3]. For a given lattice size, all possible matrices in Hermite normal form can be systematically generated by enumerating all possible factor sets and, for each factor set, iterating over all possible values of the off-diagonal elements. This presents a straightforward algorithm for assigning a unique index to any matrix in Hermite normal form for which the lattice size (the determinant of the matrix) and the dimensionality (the number of rows and columns in the matrix) are known.

We first illustrate our approach via an example. Suppose we would like to generate an index for a three-dimensional superlattice of with 15 primitive cells per supercell. We start by systematically listing all the possible permutations of ways in which 15 can be factored into three integers:

$$\begin{Bmatrix} \{1,1,15\}, \{1,3,5\}, \{1,5,3\} \\ \{1,15,1\}, \{3,1,5\}, \{3,5,1\} \\ \{5,1,3\}, \{5,3,1\}, \{15,1,1\} \end{Bmatrix} \qquad (6)$$

Each set of factors $\{f_1, f_2, f_3\}$ corresponds to the diagonal elements $\{H_{11}, H_{22}, H_{33}\}$ for the matrix in Hermite normal form. The total number of unique matrices in Hermite normal form for each set of factors is therefore given by the total number of possible combinations for the off-diagonal elements.

$$\text{number of matrices per factor set} = H_{11}^2 H_{22} = f_1^2 f_2 \qquad (7)$$

For a matrix constructed from a given factor set, we can assign a unique index from 0 to $H_{11}^2 H_{22} - 1$ based on the values of the off-diagonal elements using

$$\text{index within factor set} = H_{21} + H_{31}H_{11} + H_{32}H_{11}^2. \qquad (8)$$

The final index for the matrix is therefore

$$\text{index} = H_{21} + H_{31}H_{11} + H_{32}H_{11}^2 + \sum_{\substack{\text{Previous} \\ \text{factor sets}}} f_1^2 f_2. \qquad (9)$$

As an example, consider the following matrix in Hermite normal form, created using the $8^{\text{th}}$ factor set from those listed in equation (6).

$$\mathbf{H} = \begin{bmatrix} 5 & 0 & 0 \\ 4 & 3 & 0 \\ 0 & 2 & 1 \end{bmatrix}, \qquad (10)$$

the factorizations precede the set of its diagonal elements are

$$\begin{Bmatrix} \{1,1,15\}, \{1,3,5\}, \{1,5,3\}, \{1,15,1\} \\ \{3,1,5\}, \{3,5,1\}, \{5,1,3\} \end{Bmatrix}. \qquad (11)$$

The total number of possible HNF matrices with diagonal elements being one of these factorizations can be calculated by

$$\begin{aligned}
\text{number of matrices} &= \sum_{\substack{\text{Previous} \\ \text{factor sets}}} f_1^2 f_2 \\
&= 1^2 \times 1 + 1^2 \times 3 + 1^2 \times 5 + 1^2 \times 15 + 3^2 \times 1 + 3^2 \times 5 + 5^2 \times 1 \\
&= 1 + 3 + 5 + 15 + 9 + 45 + 25 \\
&= 103
\end{aligned} \qquad (12)$$

The number of matrices (inclusive) precede the given $\mathbf{H}$ can be calculated by

$$\text{index within factor set} = H_{21} + H_{31}H_{11} + H_{32}H_{11}^2$$
$$= 4 + 0 \times 5 + 2 \times 5^2 \qquad (13)$$
$$= 54$$

Therefore, the index for the given matrix is $103 + 54 = 157$.

---

**Algorithm 1** Assigning superlattice indices based on transformation matrices in HNF

**Input:**
    **H** - Transformation matrix in Hermite normal form (HNF).
    $n$ - Dimension of the matrix.

**Output:**
    index - Index of the superlattice.

1: **if** $n == 1$ **then**
2:     **return** $H_{11}$;
3: **else**
4:     $N_t = det(\mathbf{H})$;
5:     factorSets[][] ← Sets of factorizations of $N_t$ into $n$ integral numbers;
6:     index = 0
7:     **for** diagonals[] in factorSets[][] **do**
8:         **if** diagonals[] == $\{H_{11}, H_{22}, \ldots, H_{nn}\}$ **then**
9:             **break**;
10:         index = index + $\prod_{i=1}^{n-1}$ diagonals$[i]^{n-i}$;
11:     index = index + $\sum_{i=2}^{n} \sum_{j=1}^{i-1} [H_{ij} \cdot (\prod_{k=1}^{i-2} H_{kk}^{i-k-1}) \cdot (\prod_{k=1}^{j-1} H_{kk})]$;
12: **return** index;

Figure S4. Algorithm for assigning superlattice indices based on the HNF of transformation matrices. It is applicable to matrices with any dimensions.

Pseudocode for this process is provided in Figure S4. Elaboration on this pseudocode is provided as follows:

- This algorithm is applicable to any dimensions, not restricted to three.
- Line 5: each set of factors composes a possible set of diagonal elements of HNF matrices with determinant $N_T$. Permutations of factors are counted as different sets, since they reside on different diagonal positions of **H**. The array of factor sets is arranged in ascending order in terms of the value of each factor. For example, for $N_T = 15$, $\{1,1,15\}$ is the first factorization. $\{1,15,1\}$ precedes $\{3,1,5\}$, since the first factor of the former factorization is smaller than that of the latter.

- Line 7 – line 10: this block counts the cumulative number of matrices that can be generated by the factor sets preceding the factor sets used to construct $\mathbf{H}$ (equation (7)).
- Line 11: the second item of the right-hand side of the equation calculates the index of a matrix within a given factor set (equation (8)).

The opposite operation, which returns a transformation matrix in HNF, can be easily derived based on the same indexing logic. The pseudocode of this opposite operation is shown in Figure S5. Take the index calculated above as an example. The input index is 157 and determinant is 15. The index is within the index interval of the cumulative number of matrices for the factorization $\{5,3,1\}$. The rank of the matrix within this factorization is $157-103=54$. Then the off-diagonal elements can be calculated by

$$H_{32} = 54/(H_{11} \cdot H_{11}) = 54/25 = 2$$
$$H_{31} = 4/(H_{11}) = 4/5 = 0 \tag{14}$$
$$H_{21} = 4/1 = 4$$

The divisions are integer division and the reminder of each division is taken to calculate next off-diagonal element.

```
Algorithm 2 Regenerating the transformation matrix from an index
Input:
    n - Dimension of the matrix.
    N_T - Number of total k-points.
    index - Index of the superlattice.
Output:
    H - Transformation matrix in Hermite normal form (HNF).
Initialization:
    H - 3 × 3 zero matrix.

1:  factorSets[][] ← Sets of factorizations of N_T into n integral numbers.
2:  for diagonals[] in factorSets[][] do
3:      count = ∏_{i=1}^{n-1} diagonals[i]^{n-i};
4:      if index > count then
5:          index = index - count;
6:      else
7:          for i = 1 to n do
8:              H_{ii} = diagonals[i];                    ▷ assign diagonal elements
9:          break;
10: for i = n to 2 do
11:     for j = i - 1 to 1 do
12:         c = ∏_{k=1}^{i-2} (H_{kk}^{i-k-1}) · ∏_{k=1}^{j-1} (H_{kk})
13:         H_{ij} = index/c;                             ▷ integer division
14:         index = index%c;                              ▷ modulo operation
15: return H;
```

Figure S5. Algorithm for retrieving the superlattice from a given index.

### 3.2.2 Determination of Coefficients and Estimation of $r_{lattice}$

Ensuring that $r_{lattice} \geq r_{min}$ requires a calculation of $r_{lattice}$ for each candidate superlattice. Therefore, determination of $r_{lattice}$ by Minkowski reduction could become a major overhead. The database-query approach cuts down the computational cost by remembering which linear combination of primitive lattice vectors resulted in the shortest lattice vector every time it performs a Minkowski reduction on a candidate superlattice. The next time the same generating matrix, $H$, is encountered, the database first tries the known linear combinations of primitive vectors to see if any of them has a length less than $r_{min}$. If they do, the lattice can be eliminated from consideration without performing Minkowski reduction. If not, then full Minkowski reduction is performed. If a new linear combination of primitive vectors is found that has a length less than $r_{min}$, the coefficients of this combination are stored for future screens. In this way, the database continuously learns how to improve its performance.

The database remembers the linear combinations of primitive lattice vectors that result in a superlattice vector by projecting the superlattice vector onto a set of pre-defined mutually orthogonal vectors. For cubic, tetragonal, and orthorhombic systems, the orthogonal vectors are the conventional vectors defined in section 2.2 of the main text. For hexagonal and trigonal systems, the orthogonal vectors can be calculated by

$$\mathbf{v}_1 = \mathbf{c}_1, \mathbf{v}_3 = \mathbf{c}_3$$
$$\mathbf{v}_2 = \frac{\mathbf{c}_1 \times \mathbf{c}_3}{\|\mathbf{c}_1 \times \mathbf{c}_3\|} \cdot \|\mathbf{c}_1\|$$
(15)

where $\{\mathbf{c}_1, \mathbf{c}_2, \mathbf{c}_3\}$ are conventional lattice vectors as defined in section 2.2 of the main text, and $\{\mathbf{v}_1, \mathbf{v}_2, \mathbf{v}_3\}$ are the orthogonal vectors. Given a shortest lattice vector in a superlattice, $\mathbf{r}$, the $i^{th}$ coefficient is calculated by

$$c_i = \frac{\mathbf{r} \cdot \mathbf{v}_i}{\mathbf{v}_i \cdot \mathbf{v}_i}$$
(16)

where $\mathbf{v}_i$ is the $i^{th}$ orthogonal vector. The opposite operation, calculating the length of a stored candidate vector in a superlattice, is accomplished by

$$\|\mathbf{r}\| = \sqrt{\sum_{i=1}^{3}\left(c_i \cdot \|\mathbf{v}_i\|\right)^2}.$$
(17)

As the orthogonal vectors $\{\mathbf{v}_1, \mathbf{v}_2, \mathbf{v}_3\}$ need only be calculated once for any new query and the all sets of coefficients $\{c_1, c_2, c_3\}$ are stored in the database, equation (17) provides a rapid way to calculate an upper bound on $r_{lattice}$.

# 4. Pseudocode of Algorithm for Fast Calculation of Symmetrically Irreducible *K*-points and *K*-point Weights

---

**Algorithm 3** Fast calculation of symmetrically irreducible $k$-points and $k$-points weights

---

**Input:**
    $\{\mathbf{R}\}$ - The point symmetry group of the symmorphic space group of input structure.
    $\mathbf{H}$ - The transformation matrix of real-space superlattice in Hermite normal form.
    $N_T$ - Number of total $k$-points for the grid to be calculated for.
    $\mathbf{s}$ - Shift vector of the $\Gamma$ point.

**Output:**
    $N_i$ - Number of symmetrically irreducible $k$-point.
    cooridnates$[N_T][3]$ - Array of $k$-point cooridnates.
    weights$[N_T]$ - Array of weights.

**Initialization:**
    $\{\mathbf{R'}\}$ - Empty array of 3x3 matrices, representing symmetry operation in the basis of generating vectors, $\{\mathbf{d_1}, \mathbf{d_2}, \mathbf{d_3}\}$.
    $N_i = 0$.
    coordinates$[N_T][3]$ - Zero array of coordinates.
    weights$[N_T]$ - Zero array.

1: **for** $\mathbf{R}$ in $\{\mathbf{R}\}$ **do**
2:     $\mathbf{R'} = \mathbf{H}^T \cdot (\mathbf{R}^T)^{-1} \cdot (\mathbf{H}^T)^{-1}$;
3:     add $\mathbf{R'}$ to the array, $\{\mathbf{R'}\}$;
4: $\mathbf{J} \leftarrow$ HNF of $\mathbf{H}^T$;
5: let $\mathbf{J} = [\mathbf{j_1}, \mathbf{j_2}, \mathbf{j_3}]^T$;
6: **for** $k_3 = 0$ to $J_{33} - 1$ **do**
7:     **for** $k_2 = 0$ to $J_{22} - 1$ **do**
8:         **for** $k_1 = 0$ to $J_{11} - 1$ **do**
9:             index $= 1 + k_1 + J_{11}k_2 + J_{11}J_{22}k_3$;
10:             minIndex $\leftarrow$ initialize as $infinity$;
11:             **for** $\mathbf{R'}$ in $\{\mathbf{R'}\}$ **do**
12:                 $[k_1', k_2', k_3'] = [[k_1, k_2, k_3] + \mathbf{s}] \cdot \mathbf{R'} - \mathbf{s}$;
13:                 **if** $[k_1', k_2', k_3']$ contains non-integral values **then**
14:                     return 0, $null$, $null$;
15:                 **for** $i = 3$ to $1$ **do**
16:                     $[k_1', k_2', k_3'] = [k_1', k_2', k_3'] - \lfloor k_i'/J_{ii} \rfloor \cdot \mathbf{j_i}$;
17:                 newIndex $= 1 + k_1' + J_{11}k_2' + J_{11}J_{22}k_3'$;
18:                 minIndex $= min$(minIndex, newIndex);
19:             $[x_1, x_2, x_3] = ([k_1, k_2, k_3] + \mathbf{s}) \cdot (\mathbf{H}^T)^{-1}$;
20:             $\mathbf{k} = [x_1 \% 1, x_2 \% 1, x_3 \% 1]$;
21:             coordinates[index] $= \mathbf{k}$;
22:             **if** minIndex $==$ index **then**
23:                 $N_i = N_i + 1$;
24:                 weights[index] $= 1$;
25:             **else**
26:                 weights[minIndex] $=$ weights[minIndex] $+ 1$;
27: **return** $N_i$, cooridnates[ ][3], weights[ ];

---

Figure S6. Algorithm for fast calculation the coordinates of symmetrically irreducible *k*-points and the corresponding weights.

Line 12 – Line 14 verify whether shift vectors preserve all point symmetries. Line 16 is the reverse operation of equation (12) in the main text. in the basis of the reciprocal primitive lattice. Given a k-point, it finds the integral coordinates of the translationally equivalent k-point within the unit cell located at the origin. $\lfloor k_i' / J_{ii} \rfloor$ represents the floor operation, which returns the greatest integer no larger than the argument.

## 5. Additional Useful Algorithms

This section introduces several algorithms, which might not be essential for generating the generalized grids, but is either very useful in the process, or add extra functionalities to our server.

### 5.1 Representation of Superlattices and Identification of Symmetry-Preserving Superlattices

Each superlattice can be represented by the Hermite normal form (HNF) of the transformation matrix [3, 4], $\mathbf{M}$, where $\mathbf{M}$ is as shown in equation (7) of the main text. The Hermite normal form ($\mathbf{H}$) is defined for integral matrices and satisfies the following requirements

$$H_{ij} = 0, \quad \text{if } j > i \tag{18}$$

$$0 \leq H_{ij} < H_{jj}, \quad \text{if } j < i \tag{19}$$

Equation (18) states that a matrix in HNF is lower-triangular, and equation (19) requires that all elements are non-negative and the maximum element in each column resides on the diagonal. There is an equivalent upper-triangular formulation of the HNF of a matrix, but we use the lower-triangular one. Each non-singular integral matrix can be transformed into its Hermite normal form by multiplying a series of unimodular matrices (integral matrices whose determinant is 1 or -1). In other words, the determinant of a matrix and its HNF are equal. And it has been shown that two superlattices are equivalent if and only if the HNF of their generation matrix $\mathbf{M}$ are the same [3, 4]. An alternative, yet equivalent statement is that the HNF of a matrix is unique. The uniqueness of HNF of a matrix provides a convenient algorithm to enumerate all possible superlattices of a primitive lattice. The details have been laid out in references [3, 4]. We use a simple algorithm, developed from the uniqueness of the HNF of a matrix, to determine whether a superlattice preserves a given point symmetry operation. The algorithm works as follows:

1) Multiply the matrix representation of the point symmetry, $\mathbf{R}$, with the HNF, $\mathbf{H}$, of the transformation matrix, $\mathbf{M}$.

2) Find the HNF form of the resulting integral matrix from last step, $\mathbf{H}'$.

3) If $\mathbf{H} = \mathbf{H}'$, then this superlattice preserves this point operation $\mathbf{R}$. Otherwise, the superlattice doesn't possess this symmetry.

## 5.2 Algorithm for Detection of Structures without 3-dimensional Periodicity

In software packages that assume three-dimensional periodicity for all calculations, low-dimensional structures such as surfaces, and nanoparticles are modelled by adding vacuum to the normal directions of the periodic low-dimensional lattice. We will refer to such normal directions as the "vacuum directions". As there is little interaction between materials separated by vacuum, it is not necessary to sample more than one $k$-point in reciprocal lattice directions that are normal to the real-space periodic lattice. To ensure efficient $k$-point grids are generated in such cases, we have developed an algorithm to determine when structures are suitably separated by vacuum, and we adjust the generated $k$-point grid accordingly. For example, when simulating a slab, the density of the grid will be minimized along the direction parallel to the vacuum direction. For nanoparticles, only a single $k$-point will be returned.

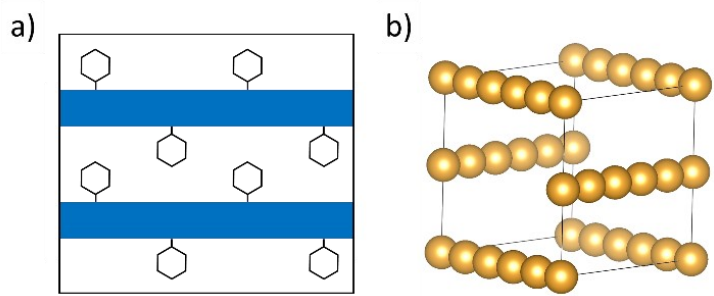

Figure S7. Two examples of low-dimensional systems recognized by our algorithm. A) A slab with adsorbed molecules. As long as the distances between all atom in one slab (including adsorbates) and the nearest atoms in a neighboring slab (including adsorbates) is at least $r_{gap}$, this will be treated as a low-dimensional system. B) An example with one-dimensional chains oriented in different directions. As long as the distances between chains are at least $r_{gap}$, the algorithm will recognize this system as being periodic in two dimensions.

The user-provided input to our algorithm is a minimum distance by which slabs, nanowires, or surfaces must be separated to trigger a change in the $k$-point grid. We call this quantity $r_{gap}$. Given a value for $r_{gap}$, our algorithm identifies gaps between slabs, nanowires, or nanoparticles regardless of the topology of the system (Figure S7). We accomplish this by starting at a single atom and recursively visiting all neighbouring atoms within a distance of $r_{gap}$. When we encounter an atom that is translationally equivalent to an atom we have already visited, we record the vector between those atoms. We refer to such vectors, which are normal to the vacuum directions, as "contiguous vectors". For a slab structure, the contiguous vectors will be parallel to the slab surface. In a nanowire, the contiguous vector is the lattice vector parallel with the nanowire. The contiguous vectors are not necessarily the input primitive lattice vectors but must be linear combinations of them. In some cases, (e.g a molecule between two slabs, or something like Figure S7 b), there may be more than one set of contiguous atoms that are separated by at least $r_{gap}$. We identify such cases by ensuring that we have visited each set of translationally equivalent atoms at least once.

**Algorithm 4** Detection of periodic sublattice in structures

**Input:**
  $x[\,]$, $\{v_i\}$ - As defined above.
  $seen[\,]$ - Array of boolean type. $seen[i]$ indicates whether the atom $x[i]$ has been passed to the function FINDCONTIGUOUSVECTORS().
  $siteIndex$ - The index of atom for which neighbouring atoms are searched.
  $precision$ - This value is used to address numerical error. Distance with an absolute value below $precision$ are treated as if they are zero.

**Output:**
  $\{v_i\}$ - Contiguous vectors.

1: **function** GETCONTIGUOUSVECTORS($x[\,]$, $seen[\,]$, $\{v_i\}$)
2:   **for** $i = 1$ to x.length **do**
3:     **if** $seen[i]$ **then**
4:       continue;
5:     **else**
6:       $seen[i] = true$;
7:       $\{v_i\}$ = FINDCONTIGUOUSVECTORS($x[\,]$, $seen[\,]$, $i$, $\{v_i\}$);
8:   **return** $\{v_i\}$;

9: **function** FINDCONTIGUOUSVECTORS($x[\,]$, $seen[\,]$, $siteIndex$, $\{v_i\}$)
10:   $y[\,] \leftarrow$ array of neighbouring atoms of $x[siteIndex]$, within a radius of $r_{gap}$;
11:   **for** $j = 1$ to y.length **do**
12:     $k \leftarrow$ index in $x[\,]$ of the translationally equivelant atom of $y[j]$;
13:     **if** not $seen[k]$ **then**
14:       $seen[k] = true$;
15:       $\{v_i\}$ = FINDCONTIGUOUSVECTORS($x[\,]$, $seen[\,]$, $k$, $\{v_i\}$);
16:     **else**
17:       $u = y[j] - x[k]$;
18:       **if** $\|u\| < precision$ **then**
19:         continue;
20:       **else if** not $parallelToLatticePlane(u, \{v_i\})$ **then**
21:         add $u$ to $\{v_i\}$;
22:     **if** v.length $== 3$ **then**
23:       **return** $\{v_i\}$;
24:   **return** $\{v_i\}$;

Figure S8. Algorithm for detection of the periodic sublattice in structures without periodicity in three dimensions.

This algorithm is shown by the pseudocode in Figure S8. Elaboration on some lines are provided as follows:

- Function FINDCONTIGUOUSVECTORS(): constructs the contiguous vectors recursively by crawling over all atoms that are separated from at least one other atom in the set by a distance no more than $r_{gap}$. When an atom that is translationally equivalent to one that has previously been visited is found, then the vector between these atoms is a candidate vector. It is added to the set of contiguous vectors if it is not spanned by the ones already in the set.
- GETCONTIGUOUSVECTORS() ensures that all atoms in the unit cell are visited by FINDCONTIGUOUSVECTORS(). This is important for situations in which there are multiple sets of contiguous atoms separated by at least $r_{gap}$ (e.g. a molecule above a slab, or something like Figure S7 b).

Once we have identified the contiguous vectors, the vacuum direction(s) are calculated as the directions that are normal to all contiguous vectors. The algorithm then distorts the input structure by stretching the primitive lattice vectors along the vacuum directions so that their projections along the vacuum directions have sizes at least $r_{min}$ ( $2r_{min}$ for nanowires and slabs). The components of the lattice vectors parallel to the contiguous vectors are not changed. This effectively tells the lattice-generation algorithm that spacing between translationally equivalent atoms is already sufficiently large in the vacuum directions, and supercells only need to be created in the directions parallel to the contiguous vectors. The grid-generation algorithm is used on the distorted structure. The coordinates of the generated *k*-points, in the basis of reciprocal lattice vectors, are the same for both the original and distorted structure. Through this approach we are able to generate low-dimensional grids that respect the symmetry of the three-dimensional calculation. The complete algorithm is summarized as pseudocode in Figure S9. Explanations for some lines are presented as follows:

- Line 3: $V_{min}$ represents the minimum supercell volume. It's equal to the volume of a primitive unit cell, $V_p$, times the minimum number of total *k*-points, $N_{min}$, that users specify.

- Line 4 – line 11: $r_{stretch}$ is the target distance by which the projections of primitive lattice vectors along the vacuum directions should be stretched to. This block demonstrates how to calculate this value for various periodicities. For n = 0, $r_{stretch}$ is equal to the larger value between $r_{min}$ and the maximum possible value of $r_{lattice}$ for a unit cell volume of $V_{min}$. The latter is achieved when the lattice is close-packed with a fcc structure. For $n=1$, line 7 gives the minimum length of the real-space superlattice vectors parallel to the one-dimensional structure. Line 13 then calculates $r_{stretch}$ by finding the larger value between $r_{min}$ and the maximum possible value of $r_{lattice}$ for a two-dimensional lattice with primitive cell area of $V_{min}/r_{periodic}$. The latter is achieved for a hexagonal lattice. For $n=2$, line 10 calculates the maximum of 1) the minimum cell area for a planar lattice for which $r_{lattice}$ is at least $r_{min}$ and 2) the cell area for the lattice formed by the found contiguous vectors. The area given by 1) can be calculated by assuming the 2-dimensional lattice is hexagonal.
- Line 13 – line 15: this code block calculates a uniform scaling ratio for all lattice vectors. The maximum ratio is selected to ensure the projections of all lattice vectors along vacuum directions have a length at least $r_{stretch}$.

**Algorithm 5** Adjusting structures without three-dimensional periodicity

**Input:**
    x[ ] - Array of atomic coordinates in the basis of $\{a'_1, a'_2, a'_3\}$.
    $r_{gap}$ - Minimum gap size between atoms, larger than which a gap is considered as a vacuum.
    $N_{min}$ - The minimum number of total $k$-points specified by users.
    $\{a_1, a_2, a_3\}$ - Primitive lattice vectors.
    $V_p$ - Volume of the primitive unit cell.

**Output:**
    $\{a'_1, a'_2, a'_3\}$ - Stretched lattice vectors whose projections along the vacuum directions having
        sizes at least the $2 \times r_{min}$.

**Initialization:**
    $\{v_i\}$ - An empty array of contiguous vectors.
    seen[ ] - An array of boolean type with x.length elements initialized as *false*.

1: $\{v_i\}$ = GETCONTIGUOUSVECTORS(x[ ], seen[ ], $\{v_i\}$);
2: $n \leftarrow$ dimensions of $\{v_i\}$;
3: $V_{min} = V_p \times N_{min}$;
4: **if** $n == 0$ **then**
5:     $r_{stretch} = max(r_{min}, \sqrt[3]{V_{min}/\sqrt{2}})$;
6: **else if** $n == 1$ **then**
7:     $r_{periodic} = max(r_{min}, \|v\|)$;
8:     $r_{stretch} = 2 \times max(r_{min}, \sqrt{(2\sqrt{3})/3 \cdot V_{min}/r_{periodic}})$;
9: **else if** $n == 2$ **then**
10:     $S_{periodic} = max(\sqrt{3}/2 \cdot r_{min}^2, \|v_1 \times v_2\|)$;
11:     $r_{stretch} = 2 \times max(r_{min}, V_{min}/S_{periodic})$;
12: scalingRatio = 0;
13: **for** $a_i$ in $\{a_1, a_2, a_3\}$ **do**
14:     $n_i \leftarrow$ projection of $a_i$ to the normal directions of lattice $\{v_i\}$;
15:     scalingRatio = $max$(scalingRatio, $r_{stretch}/\|n_i\|$);
16: $\{e_i\} \leftarrow$ unit vectors normal to $\{v_i\}$;
17: **for** $a_i$ in $\{a_i\}$ **do**
18:     **for** $e_j$ in $\{e_i\}$ **do**
19:         $a'_i = a_i + [(\text{scalingRatio} - 1) \cdot (a_i \cdot e_j)] \cdot e_j$;
20: **return** $\{a'_1, a'_2, a'_3\}$;

Figure S9. Algorithm for stretching lattice vectors to reduce *k*-point density along the vacuum directions accordingly.

## 6. Implementations

In addition to the kpLib library, which helps integrate the generalized grids into simulation packages, two more implementations are provided to meet the diverse demands of users. The server was initially launched in our previous work, but extensive updates has been performed since then, both to increase robustness and to improve the database. Last year especially, the total number of grids in the latest version of database was increased up by 637%. The server is also wrapped into a stand-alone application, which is portable across different platforms and is desirable for scientific computing clusters without internet access.

### 6.1 *K*-Point Grid Server: A Ready-to-use Online Application

The *K*-Point Grid Server, referred as "the server" below, is a ready-to-use internet-based application. It generates the optimal generalized Monkhorst-Pack grids by dynamic grid generation for triclinic and monoclinic systems, and by rapidly searching a pre-generated database, as discussed in section 2, for all other crystal systems. The database contains generalized *k*-point grids calculated from all symmetry-preserving superlattices of a set of 16,808 sample structures in cubic, hexagonal, trigonal, tetragonal, and orthorhombic crystal systems with different lattice parameters. The ratio of the longest conventional lattice vector to the shortest one is up to 64. Such dense sampling of the possible lattice parameters should make the database comprehensive enough to cover nearly all input structures from users. The maximum sizes of the superlattices are 46,656 (36×36×36) for cubic and 5832 (18×18×18) for triclinic, monoclinic, cubic and the other four crystal systems, the same as the search depths discussed in section 2 of the SI. The scale factor is used when requests exceed these grid sizes. The database searching approach saves the computational cost of enumerating the superlattices and counting the symmetrically distinct *k*-points in corresponding grids for every user request, which gives it an advantage over dynamic grid generation. For monoclinic and triclinic systems, however, the server uses the dynamic searching scheme, since the database searching approach wouldn't be as beneficial as it is to the other five Bravais lattices because of the huge number symmetry-preserving superlattices for these two systems. The algorithm for detecting vacuum spaces and correspondingly adjusting the *k*-point grid is also implemented in the server, as are other algorithms for determining symmetry that are specific to the *ab-initio* software package being used.

Users can tailor their requests to the server through a set of parameters defined in a file named "PRECALC", and the server is queried using a small script called "getKPoints". Grid sizes are specified through either MINDISTANCE or MINTOTALKPOINTS, which correspond to $r_{min}$ and $N_{min}$ respectively. An example of a PRECALC file, the getKPoints script, and a detailed description of all allowed parameters in PRECALC can be found on our website (http://muellergroup.jhu.edu/K-Points.html).

**6.2 *K*-Point Grid Generator: An Open-source, Stand-alone Application**

The *K*-Point Grid Generator is a self-contained application for users with runtime environments that might not have an internet connection. It has the exact same set of functionalities as the server and is updated accordingly every time a new version the server is released. In addition, the dynamic generation method is implemented, and is used to generate grids for monoclinic and triclinic systems, which are not covered by database because of the large number of entries there would be. Users still specify parameters through a PRECALC file and launch the application through a script getKPoints. But the script is different from the one used for server and is tailored for the stand-alone application. The Java programming language is used to ensure the portability and a consistent performance across operating systems. The project is open sourced through a public repository (https://gitlab.com/muellergroup/k-pointGridGenerator). A pre-built binary can be downloaded from our website and is packaged with the tailor calling script getKPoints and with a complete set of files for the database. The database files are stored in binary, gzipped format and take up about 7.15 MB of disk space.

## 7. Additional Benchmarks

### 7.1 Speed Comparison between Database Lookup and Dynamic Generation

Additional speed benchmarks were performed on MINTOTALKPOINTS ($N_{min}$), representing another common way to specify grid sizes in input files of our tools. They were also conducted for both $\Gamma$-centered grids and grids that for which the best shift vectors were automatically selected. The latter grids are referred as "auto grids" in the following. For each grid generation, three generation schemes were used: the database lookup by the stand-alone application, the dynamic search by the kpLib library and the dynamic search by kpLib library with scale factor turned on. Since the database only contains grids for the five crystal systems excluding triclinic and monoclinic ones, we performed benchmarks using the 87 out of 102 structures (the same set used for benchmarking in the main text) belonging to those five crystal systems. Each calculation was repeated three times and the average computation time was taken as the time for grid generation. The time measurement only takes in to account the actual grid generation time and excludes the time spent for initialization and input/output operations. All benchmarks were performed on the Homewood High-Performance Cluster (HHPC) using Intel Xeon E5660 processors with a 2.80 GHz base frequency and a 48 GB RAM.

Figure S10 shows grid-generation times based on user-specified values of $N_{min}$ ranging from 1 to 31,622. These numbers were picked randomly to give a relatively even sampling of $N_{min}$ when plotted on a logarithmic scale. The database lookup is the fastest method. The usage of the scale factor significantly reduces the computation time for large grids. However, there are some noticeable differences with the benchmarks based on user-provided values for $r_{min}$ (section 4.1 of main text). First, generating grids based on $N_{min}$ takes more time on average than using $r_{min}$. In addition, the dynamic generation of $\Gamma$-centered grids, although it has no shift vectors to iterate over, is computationally more costly. For example, the auto grids generation at $N_{min} = 7943$ complete at 6.61 seconds, while it take about 163.87 seconds for the $\Gamma$-centered case. The substantial difference in speed is because with shift vectors, the algorithm is more likely to find nearly optimal grids early in the search. Therefore, $N_{upper}$ for dynamic generation for an auto grid is much smaller than that for $\Gamma$-centered grid generation.

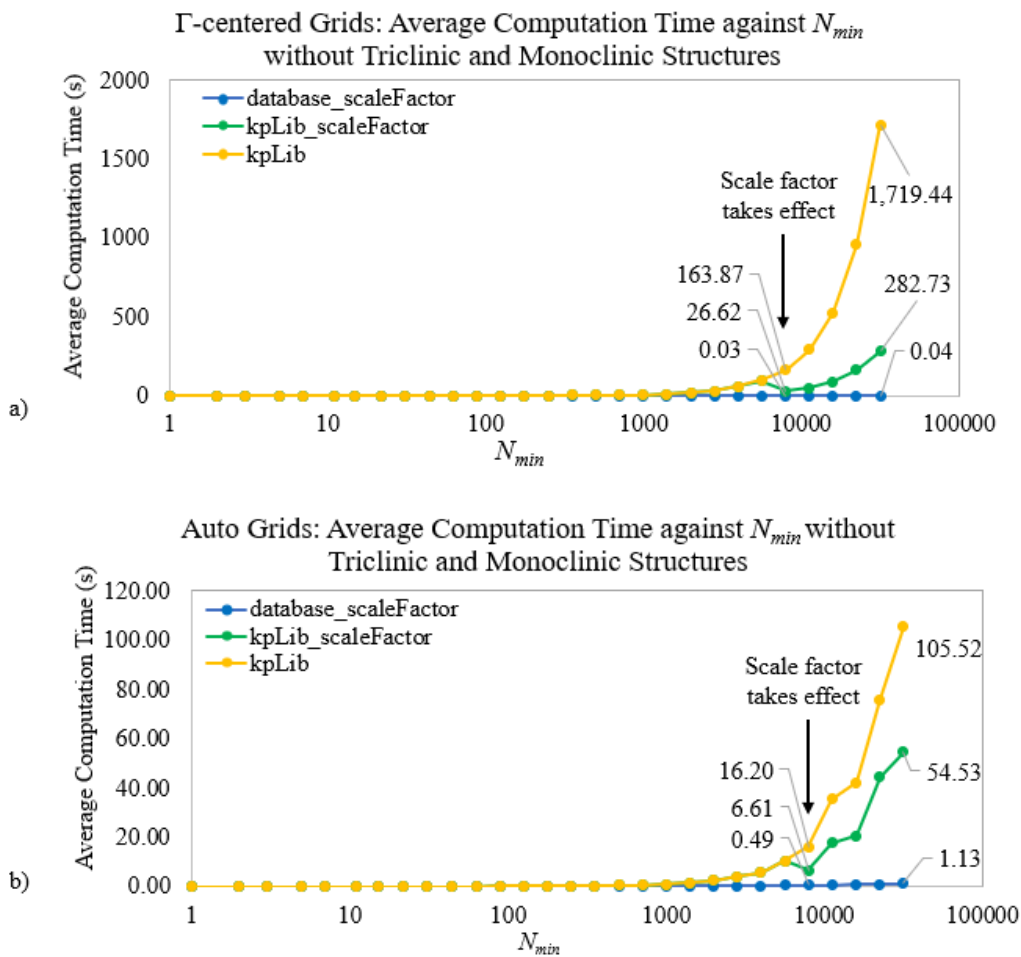

Figure S10. Average computation time of three grid generation methods over randomly selected structures without triclinic and monoclinic ones at $N_{min}$ ranging from 1 to 31,622 for a) Γ-centered grids, b) auto grids. The smallest $N_{min}$ at which the scale factor starts to take effect is 7,943. The longest computation times and the times at the value where scale factor takes effect are labeled in the graphs.

## 7.2 Acceleration with $r_{min}$ Being the Limiting Factor

The algorithms for enumerating symmetry-preserving superlattices in section 2.3 of the main text can be accelerated by enforcing $r_{lattice} > r_{min}$ at each step of constructing a superlattice. This allows the algorithm to skip many superlattices at an early stage. To measure the degree of acceleration by screening based on $r_{min}$, we benchmarked the computation time for generating a generalized k-point grid for both Γ-centered and shifted grids, with the demonstration application in C++ using kpLib. In each case, generalized *k*-point grids were generated at three values of $r_{min}$: 25, 50, and 75 angstroms. Each calculation was repeated five times and the average response time was recorded as the calculation time for that structure. The computation time for each crystal system is taken as the average time of structures within the 102 materials that belongs to this system. The benchmark was performed on a virtual machine with Ubuntu 18.04 operation system, and on Intel Core i7-8550U processors with a clock speed of 1.80GHz. The ratios of the computation time between non-accelerated and accelerated codes are plotted in Figure S11. Results demonstrate a significant acceleration, and the amount of acceleration increases as $r_{min}$ grows. Consistent acceleration is observed for both the Γ-centered and shifted grids. The highest ratio is 37.4 for Γ-centered grids with $r_{min}$ equal to 75 angstroms in trigonal system, and is 32.8 for shifted grids with $r_{min}$ of 75 angstroms in cubic system. The average ratio across all seven systems at 75 angstroms are 21.5 for Γ-centered grids and 16.2 for shifted grids.

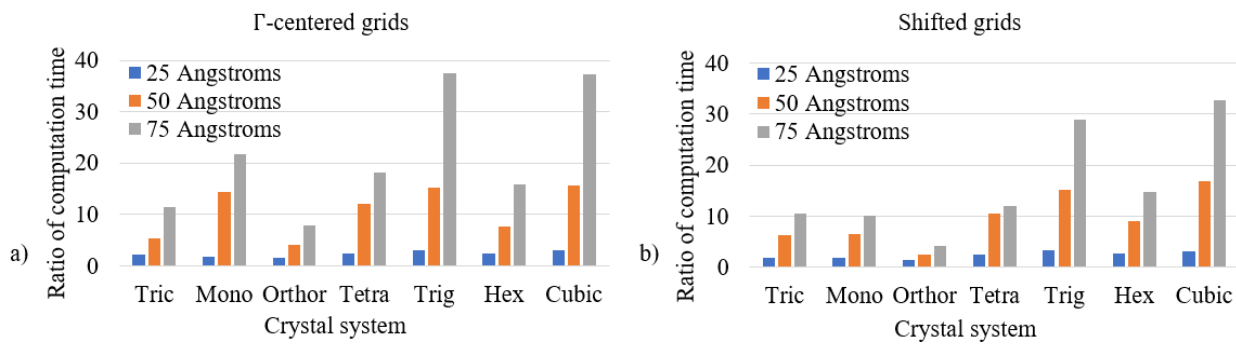

Figure S11. Ratios of computation time between the non-accelerated and accelerated algorithm for all seven crystal systems with $r_{min}$ at 25, 50, and 75 angstroms, for a) Γ-centered grids and b) shifted grids. The *x*-axis lists the crystal systems. From left to right, they represent triclinic, monoclinic, orthorhombic, tetragonal, trigonal, hexagonal and cubic, respectively.

# References

1. Wisesa, P., K.A. McGill, and T. Mueller, *Efficient generation of generalized Monkhorst-Pack grids through the use of informatics.* Physical Review B, 2016. **93**(15).
2. Giacovazzo, C., et al., *Fundamentals of Crystallography*. third ed. IUCr Texts on Crystallography. 2002, Oxford: Oxford University Press. 842.
3. Mueller, T., *Computational studies of hydrogen storage materials and the development of related methods*, in *Department of Materials Science and Engineering*. 2007, Massachusetts Institute of Technology: Boston, Massachusetts. p. 127-132.
4. Hart, G.L.W. and R.W. Forcade, *Algorithm for generating derivative structures.* Physical Review B, 2008. **77**(22).